\documentclass[12pt]{iopart}
\usepackage{amssymb}
\usepackage{graphicx,bm, subeqn}
\usepackage{setstack, cite}
\begin{document}

\title[]{Ladder operators and squeezed coherent states of a 3-dimensional
generalized isotonic nonlinear oscillator}

\author{V~Chithiika Ruby, S~Karthiga and  M~Senthilvelan}

\address{Centre for Nonlinear Dynamics, School of Physics,
Bharathidasan University, Tiruchirappalli - 620 024, India.}
\ead{velan@cnld.bdu.ac.in}

\begin{abstract}
We explore squeezed coherent states of a $3$-dimensional generalized isotonic
oscillator whose radial part is the newly introduced generalized isotonic oscillator whose
bound state solutions have been shown to admit the recently discovered $X_1$-Laguerre
polynomials. We construct a complete set
of squeezed coherent states of this oscillator
by exploring the squeezed coherent states of the radial part and combining the latter with the squeezed
coherent states of the angular part. We also prove that the three mode squeezed coherent
states resolve the identity operator. We evaluate Mandel's $Q$-parameter of the
obtained states and demonstrate that these states exhibit sub-Possionian and super-Possionian photon statistics.
Further, we illustrate the squeezing properties of these states,
both in the radial and angular parts, by choosing appropriate observables in
the respective parts. We also evaluate Wigner function of these three mode squeezed coherent states and
demonstrate squeezing property explicitly.
\end{abstract}

\pacs{03.65.-w, 03.65.Ge, 03.65.Fd}

\maketitle

\section{Introduction}
\subsection{New exactly solvable model}
Very recently the exact quantum solvability of the extended radial oscillator/ generalized isotonic
oscillator potential,
\begin{eqnarray}
V(r) = \omega^2 r^2 + \frac{l(l+1)}{r^2} + \frac{8 \omega}{(2\omega r^2 + 2 l + 1)}
       - \frac{16 \omega (2l+1)}{(2\omega r^2 + 2 l + 1)^2},
\label{poten}
\end{eqnarray}
where $\omega$ and $l$ are parameters, has been demonstrated in four different perspectives \cite{ques, hall1, hall2, new1}.
To begin with, it was shown that the bound state eigenfunctions of
the Schr\"{o}dinger equation associated with the potential (\ref{poten})
can be expressed in terms of the newly found exceptional orthogonal ${\nu}^{th}$ degree polynomials,
$\hat{L}^{k}_{\nu}(x), \; \nu = 0, 1, 2,3, ...$ where $k$ is a positive real parameter,
namely $X_{1}$-Laguerre polynomials
\cite{ques}.  These new polynomials, $\hat{L}^{k}_{\nu}(x)$, are related to the classical
Laguerre polynomials $L^{k}_{\nu}(x)$ by the following relation \cite{kam}
\begin{eqnarray}
\qquad \qquad \hat{L}^{k}_{\nu}(x)=-(x+k+1){L}^{k}_{\nu-1}(x)+{L}^{k}_{\nu-2}(x).
\label{exc}
\end{eqnarray}

The bound state solutions of the Schr\"{o}dinger equation associated with
the potential (\ref{poten}) are found to be \cite{ques}
\begin{eqnarray}
\fl \qquad \qquad \qquad \qquad \phi _{n, l}(r)&=&N_{n, l}\frac{r^{l+1}}{(2 \omega r^2 + 2 l + 1)}  \hat{L}^{\left(l+\frac{1}{2}\right)}_{n+1}(\omega r^{2})\;
e^{-\frac{1}{2}\omega r^2},
\label{efn}
\end{eqnarray}
with
\begin{eqnarray}
\fl \qquad \qquad \qquad \qquad \quad E_{n, l} &=& 2 \omega \left(2 n+ l+ \frac{3}{2}\right), \qquad n=0,1,2,3...,\;l = 0, 1, 2, 3, ...
\label{sol}
\end{eqnarray}
and  ${\displaystyle N_{n, l} = \left(\frac{8\; \omega^{l + \frac{3}{2}} n!}{
(n+l+\frac{3}{2}) \Gamma(n+l+\frac{1}{2})}\right)^{1/2}}$ is the normalization constant.
We note here that the $X_1$-Laguerre polynomials, $\hat{L}^{k}_{\nu}(x)$, are solutions of the
following second-order linear ordinary differential equation with rational
coefficients, that is
\begin{eqnarray}
y'' - \frac{(x - k)(x + k + 1)}{x(x+k)} y' + \frac{1}{x}\left(\frac{x-k}{x+k} + \nu - 1\right) y = 0, \;\;\; ' = \frac{d}{dx},
\label{diff}
\end{eqnarray}
where $k > 0$ is a real parameter and $\nu = 1, 2, 3, ...$ .  The first few
$X_1$-Laguerre polynomials are \cite{kam}
\begin{eqnarray}
\hat{L}^{k}_1(x) &=& -(x+k+1), \\
\hat{L}^{k}_2(x) &=& x^2-k(k+2), \\
\hat{L}^{k}_3(x) &=& -\frac{1}{2}x^3 + \frac{k+3}{2} x^2 + \frac{k(k+3)}{2} x - \frac{k}{2}(3 + 4k + k^2).
\label{x1l}
\end{eqnarray}
One may observe that the above polynomial sequence starts with a linear
polynomial in $x$ instead of a constant which usually the other classical orthogonal polynomials do.
However, these new polynomials form complete set with respect to some
possible measure \cite{kam}. For more details on the properties of these exceptional orthogonal polynomials,
one may refer Ref. \cite{kam}.

In a different context, while constructing exact analytic solutions
of the $d$-dimensional Schr\"{o}dinger equation associated with the generalized
quantum isotonic nonlinear oscillator potential (\ref{poten}), that is
\begin{eqnarray}
\qquad \quad -\Delta  \Psi + \left(\frac{B^2}{r^2} + \omega^2 r^2 + 2 g \frac{(r^2 - a^2)}{(r^2 + a^2)^2}\right)\Psi = E\Psi,
\label{ddim}
\end{eqnarray}
where $\Delta$ is the $d$-dimensional ($d \ge 2$) Laplacian operator and
 $\omega, a, g$ are parameters and $V(r)$ is the central potential given
by
\begin{eqnarray}
V(r) = \frac{B^2}{r^2} + \omega^2 r^2 + 2 g \frac{(r^2 - a^2)}{(r^2 + a^2)^2}, \quad  B^2 \ge 0,
\label{poten2}
\end{eqnarray}
Hall et al have shown that only for the values $g = 2$ and $\omega^2 a^4 = B^2 + \left( l + \frac{(d-2)}{2}\right)^2$
the potential (\ref{poten2}) can be exactly solvable  \cite{hall1}.  It is shown that the potential
(\ref{poten2}) can be regarded as a supersymmetric partner of the
Goldmann - Krivchenkov potential whose exact solutions are known \cite{hall1}.
The potential (\ref{poten2})  with the parametric restrictions given above
exactly matches with the one constructed by Quesne \cite{ques}. In a subsequent study
\cite{hall2}, Saad et al have solved the system (\ref{poten2})
numerically, for arbitrary values of the system parameters,  through
asymptotic iteration method. They have shown that for
certain restrictions ($B = 0$ and $d = 1$) the potential (\ref{poten2})
can be reduced to the one discussed in Ref. \cite{car} (see equation (\ref{pot1}) below
and the paragraph that follows).  The quasi polynomial solutions
and the energy eigenvalues of (\ref{poten2}) are also reported in Ref. \cite{hall2}.

In a very recent paper, Agboola and Zhang have considered the potential (\ref{poten})
and transformed the associated  Schr\"{o}dinger equation
into the generalized spheroidal wave equation \cite{gswe}.  Using Bethe ansatz method,  they have derived
bound state solutions of (\ref{poten}) for certain parametric values \cite{new1}.

We mention here that the inverse square type potentials play an important role in physics \cite{bos, bec, fer, cos, aho}.
Solving Schr\"{o}dinger equation with non-polynomial potentials \cite{non} is also of interest for physicists at both
the classical \cite{iso} and  quantum level \cite{car1}.

We also recall here that
during the past few years the one dimensional version of
the potential (\ref{poten}), that is
\begin{eqnarray}
\qquad \quad V(x) = \omega^2 x^2 + 2 g\frac{ (x^2 - a^2)}{(x^2 + a^2)^2}, \quad a^2 > 0,
\label{pot1}
\end{eqnarray}
is studied in different perspectives, see for example
Refs.  \cite{hall2, car, fellows, berger, sesma, sen, chi1, chi2}.
The generalized isotonic oscillator potential (\ref{pot1})
is generalized in such a way that it
lies between the harmonic oscillator and isotonic oscillator potentials \cite{car}.
Importantly, this generalization removes the singularity nature of the isotonic oscillator
in real space, as the generalized isotonic oscillator has poles only
at imaginary points $x = \pm i a$.  The energy values of this generalized isotonic
oscillator potential $(g = 2 \omega a^2 (1 + 2 \omega a^2),\;\omega = 1,\;a^2 = \frac{1}{2})$
are equidistant \cite{car}.
In our earlier studies, we have constructed different types of nonlinear coherent states,
nonlinear squeezed states  and various nonclassical states of the oscillator (\ref{pot1})
and analyzed its position dependent mass Schr\"{o}dinger equation
by fixing the parameters $ g = 2 \omega a^2 (1 + 2 \omega a^2),\;\omega = 1$ and $a^2 = \frac{1}{2}$
\cite{chi1, chi2, chi3, chi4}.  While constructing nonlinear coherent and squeezed states
of the system (\ref{pot1}), the deformed ladder operators
which we had constructed lead to two non-unitary displacement
operators  and two squeezing operators \cite{chi2, chi3, chi4}. While one of the displacement operators
produced nonlinear coherent states  the  other non-unitary displacement
operator failed to produce any new type of nonlinear coherent states (dual pair) \cite{dual}. In the case of squeezed states also,
while one of the squeezing operators gave the nonlinear squeezed states the other one failed to produce
the dual pair of nonlinear squeezed states which lead us to the conclusion that
the dual pair of nonlinear coherent and squeezed states are absent in this system
(for more details one may refer the Refs. \cite{chi3, chi4}).

\subsection{Present work}
As a continuation of our earlier studies, in this paper, we focus
our attention on the three dimensional extension of the potential
(\ref{pot1}).  With the restriction $d = 3$, $g = 2, B = 0$ and $\omega a^2 = l + \frac{1}{2}$, the
Schr\"{o}dinger equation (\ref{ddim}) turns out to be an exactly solvable three dimensional isotonic oscillator
equation, that is
\begin{eqnarray}
-\Delta  \Psi + \left(\omega^2 r^2 + \frac{8\omega}{2\omega r^2+2l+1}-\frac{16\omega(2l+1)}{(2\omega r^2+2l+1)^{2}}\right)
 \Psi = E \Psi.
\label{3_dim}
\end{eqnarray}
The isotonic oscillator with the
generalized inverse type of potential models (i) the dynamics of harmonic oscillator in the
presence of dipoles where it describes the scattering off electrons by polar molecules \cite{pol}
and (ii) the dynamics of dipoles in Cosmic strings \cite{cos}.

Now considering the three dimensional Laplacian in polar coordinates with the wavefunction
$\Psi(r, \theta, \phi) = u(r)\;Y_{l, m}(\theta, \phi)$,
where $Y_{l, m}(\theta, \phi)$ are the spherical harmonics, we have the Schr\"{o}dinger equation
\begin{eqnarray}
\fl \;\;-\frac{d^{2}u}{dr^{2}}-\frac{2}{r}\frac{du}{dr}+\left(\omega^2 r^2+\frac{l(l+1)}{r^2}+\frac{8\omega}{2\omega r^2+2l+1}-\frac{16\omega(2l+1)}{(2\omega r^2+2l+1)^{2}}\right)u = E u,
\label{3dim}
\end{eqnarray}
where $l$ is the angular momentum.

Since the bound state solutions of (\ref{3dim}) are already known (vide equations
(\ref{efn}) and (\ref{sol})) we can unambiguously write the eigenfunctions and energy values
of the {\it three dimensional problem} as \cite{ques, hall1}
\begin{eqnarray}
\fl \quad \quad \Psi_{n, l} (r,\theta ,\phi  ) = N_{n, l} \frac{r^l}{2 \omega r^2 + 2l +1} \hat{L}^{\left(l+\frac{1}{2}\right)}_{n+1}(\omega r^2) e^{-\frac{1}{2} \omega r^2} Y_{l,m}(\theta ,\phi ),\\
\nonumber & &\\
\fl \qquad \qquad E_{n,l} = 2\omega \left(2 n+l+\frac{3}{2}\right),\; n= 0,1,2,3,..., \; l= 0,1,2,3,..., \; m = -l\; \mbox{to}\; l.\;
\end{eqnarray}

From these solutions we construct a complete set of squeezed coherent states for
the three dimensional system (\ref{3_dim}).  Since the potential under
consideration is a spherically symmetric one, we separate radial part from
the angular part and investigate each one of them separately.
To construct squeezed coherent states of the radial part, vide
equation (\ref{poten}), we should know the ladder operators of it.
The ladder operators can be explored through supersymmetric technique.  However, the resultant
operators which come out through this formalism do not  perform perfect annihilation and
creation actions on the number states.  To explore the perfect ladder operators we alternatively
examine  shape invariance property of the
potential (\ref{poten}). Once the presence of this property is confirmed, we then move on to  construct
the necessary annihilation and creation operators
by adopting the procedure described in Ref. \cite{bal}.  Using these operators we
unambiguously construct squeezed coherent states \cite{scs, obada} of the radial part.
To explore squeezed coherent states of the angular part we use Schwinger representation and  define
the necessary creation and annihilation operators. In particular, by  considering disentangled form of two
mode squeezing and displacement operators  we  derive squeezed coherent states \cite{abir}. We obtain
a complete set of squeezed coherent states of the $3$-dimensional oscillator by defining these states are
the tensor product between squeezed coherent states of the radial variable and squeezed states
of the angular variables \cite{nito}. To make our result more rigorous
we also prove that these  three mode squeezed coherent states satisfy the
completeness condition.  We also evaluate Mandel's $Q$-parameter of the three mode
squeezed coherent states and investigate certain photon statistical properties
associated with these states. Our result  confirms that these states exhibit sub-Poissonian
(non-classical) and super-Poissonian
photon statistics for certain parameters. Further,
we demonstrate squeezing properties of the states associated with the radial part  by
considering the generalized position and momentum coordinates with the obtained ladder operators
and  the angular part by considering the angular momentum quantities $\hat{L}_{x}$ and $\hat{L}_{y}$
respectively. Finally, we  evaluate the  Wigner function of the constructed squeezed coherent states by defining a three-mode Wigner
function in terms of coherent states \cite{spara, spara_N}. Our result reveals that
these squeezed coherent states exhibit the squeezing property. We intend to carry out  all these
studies due to the fact that  among the nonclassical states squeezed states have attracted much attention because
certain observables in this basis show the fluctuations less than that
of the vacuum.  The squeezed coherent states have found applications not only in the
quantum optics but also in quantum cryptography \cite{cry},  quantum teleportation
\cite{inf} and quantum communication \cite{comm}, to name a few.

This paper is organized as follows.  In  section 2, we analyze the shape invariance
property of the radial part of the $3$-dimensional generalized isotonic oscillator and construct
suitable ladder operators.  In section 3, we construct squeezed coherent states of the
three dimensional oscillator.  We use these ladder operators
to construct squeezed coherent states of the radial part and Schwinger's operators to
construct angular part.  We express the complete form of the squeezed coherent states of the
$3$-dimensional oscillator by suitably combining  the radial part with
the angular part. We also prove that the three mode
squeezed coherent states satisfy the completeness condition. We investigate certain non-classical
properties associated with these states in section 4. To begin with, we study
Mandel's $Q$-parameter and confirm the non-classical and super-Poissonian nature of these
states. We also illustrate the squeezing properties of the
constructed $3$-dimensional squeezed coherent states both in
the radial part and in the angular part in this section. In section 5, we evaluate
Wigner function for the $3$-dimensional generalized isotonic oscillator
and demonstrate the squeezing property of the constructed states.  Finally, in section 6,
we present the conclusion and the outcome of our study.  Certain background derivations
which are needed to obtain the necessary ladder operators are given in the Appendix A. The
details of evaluating certain multiple sums in the expressions $\langle \hat{n}_r \rangle$ and $\langle \hat{n}^2_r \rangle$ are
given in Appendix B. In Appendix C we discuss the method of evaluating the multiple sums appearing in the
Wigner function.

\section{Shape invariance property and ladder operators}
\subsection{Supersymmetric formalism}

To construct  annihilation and creation operators,
we start with time-independent  Schr\"{o}dinger
equation,
\begin{eqnarray}
\hat{H} \Phi(r) = \left(-\frac{d^2}{d r^2} + V(r) \right) \Phi(r) = E \Phi(r),
\label{one}
\end{eqnarray}
with $V(r)$ is given in (\ref{poten}).  Let us rewrite the
second order differential operator $\hat{H}$ given in (\ref{one})
as a product of two first order differential operators, namely $\hat{A}^{+}$ and
$\hat{A}^{-}$, such that $\hat{H}^{(1)} = \hat{H} -\omega(2l+3) =  \hat{A}^{+} \hat{A}^{-}$ and
\begin{eqnarray}
\fl \qquad \qquad \qquad \quad \hat{A}^{+ } = - \frac{d}{dr} + W(r)\;\;\;\mbox{and} \;\;\; \hat{A}^{-}=  \frac{d}{dr} + W(r).
\label{ops}
\end{eqnarray}
In the above $W(r)$ is the superpotential which is found to be \cite{ques}
\begin{eqnarray}
\quad W(r)=\omega r-\frac{l+1}{r}+ \frac{4\omega r}{2\omega r^2+2l+1}-\frac{4\omega r}{2\omega r^2+2l+3}.
\end{eqnarray}

The operators $\hat{A}^{+}$ and $\hat{A}^{-}$ factorize the partner Hamiltonian $\hat{H}^{(2)}$ as
\begin{eqnarray}
\fl \hspace{5cm} \hat{H}^{(2)} = \hat{A}^{-} \hat{A}^{+} = \hat{H}- \omega(2l+1). \nonumber
\label{part}
\end{eqnarray}

The Schr\"{o}dinger equation associated with $\hat{H}^{(1)}$ reads
\begin{eqnarray}
 \qquad \qquad \hat{H}^{(1)}\;\Phi_{n, l}^{(1)}(r)=E^{(1)}_{n}\;\Phi_{n, l}^{(1)}(r),
 \label{hama}
\end{eqnarray}
that is
\begin{eqnarray}
\fl\qquad \qquad \qquad -\frac{d^2\Phi_{n, l}^{(1)}(r)}{dr^2}+V_{1}(r)\Phi_{n, l}^{(1)}(r)=E_{n}^{(1)}\Phi_{n, l}^{(1)}(r),
\label{sys1}
\end{eqnarray}
where the potential ${V_1 (r)}$ is given by
\begin{eqnarray}
\fl \qquad \qquad V_1(r)=\omega^2 r^2+\frac{l(l+1)}{r^2}+\frac{8\omega}{2 \omega r^2+2l+1}-\frac{16\omega(2l+1)}{(2\omega r^2+2l+1)^2} - \omega (2 l + 3).
\label{v1}
\end{eqnarray}
Equation (\ref{sys1}) shares the same solution as that of the Schr\"{o}dinger equation associated
with the potential (\ref{poten}) admits, that is
\begin{eqnarray}
{\Phi_{n,l}^{(1)}\equiv \Phi_{n,l}} = N_{n, l}\frac{r^{l+1}}{(2 \omega r^2 + 2 l + 1)}  \hat{L}^{\left(l+\frac{1}{2}\right)}_{n+1}(\omega r^{2})\;e^{-\frac{1}{2}\omega r^2}
\label{efn1}
\end{eqnarray}
with the energy eigenvalues
\begin{eqnarray}
\qquad {E_{n}^{(1)}=E_{n, l}-\omega (2l+3)=4n\omega}.
\label{en1}
\end{eqnarray}

The Schr\"{o}dinger equation associated with $\hat{H}^{(2)}$ is  given by
\begin{eqnarray}
\qquad \qquad \quad \quad \hat{H}^{(2)} \Phi_{n,l}^{(2)}(r)= E_{n}^{(2)}\Phi_{n,l}^{(2)}(r).
\label{ham1}
\end{eqnarray}
Rewriting (\ref{ham1}), we get
\begin{eqnarray}
\fl \qquad \qquad \; \quad -\frac{d^2\Phi_{n,l}^{(2)}(r)}{dr^2}+V_2(r)\;\Phi_{n,l}^{(2)}(r)=E_{n}^{(2)}\Phi_{n,l}^{(2)}(r),
\label{sys2}
\end{eqnarray}
where the partner potential $V_{2}(r)$  turns out to be
\begin{eqnarray}
\fl \qquad V_2(r)=\omega^2 r^2+\frac{(l+1)(l+2)}{r^2}+\frac{8\omega}{2\omega r^2+2l+3}-\frac{16 \omega (2l+3)}{(2\omega r^2+2l+3)^2} - \omega (2 l + 1).
\label{v2}
\end{eqnarray}

The only difference between the equations (\ref{v1}) and (\ref{v2}) is that in the latter
the parameter $l$ is modified as $l+1$ besides the constant term.
As a consequence, from the known solutions, (\ref{efn}) and (\ref{sol}),
we can generate the solutions of the system (\ref{sys2}) straightforwardly, that is
\begin{eqnarray}
 \qquad  \quad  \Phi_{n,l}^{(2)}(r) = \tilde{N}_{n, l}\frac{r^{l+2}}{2\omega r^2+2l+3}\hat{L}_{n+1}^{\left(l+\frac{3}{2}\right)}(\omega r^2)\;
                                  e^{-\frac {1}{2} \omega r^2},
\end{eqnarray}
with
\begin{eqnarray}
 \qquad \quad  E_{n}^{(2)} = E_{n,l+1} - \omega(2l+1) = 4\omega(n+1)
\end{eqnarray}
and ${\displaystyle \tilde{N}_{n, l} = \left(\frac{8\;\omega^{l + \frac{5}{2}} n!}{
(n+l+\frac{5}{2}) \Gamma(n+l+\frac{3}{2})}\right)^{1/2} = N_{n, l+1}}$ is the normalization constant.

Now we investigate the action of SUSY operators, $\hat{A}^{\pm}$, on the eigenfunctions
$\Phi^{(1)}_{n, l}(r)$  with the help of  (\ref{efn1}). For this purpose let us calculate the action of
$\hat{A}^{+}$ on $\Phi_{n, l+1}^{(1)}$, that is
\begin{eqnarray}
\fl \qquad \quad \quad \hat{A}^{+} \Phi^{(1)}_{n, l+1} &=& \left(-\frac{d}{dr} + W(r)\right)
                          \frac{N_{n, l+1}\; r^{l+2}\; \hat{L}^{\left(l+\frac{3}{2}\right)}_{n+1}}{(2\omega r^2 + 2 l + 3)}e^{-\frac{1}{2}\omega r^2}. \label{ap}
\end{eqnarray}
Evaluating the right hand side of (\ref{ap}) we find
\begin{eqnarray}
 \fl \qquad  \quad \quad   \hat{A}^{+} \Phi^{(1)}_{n, l+1}    &=& \left[-\frac{d \hat{L}^{\left(l+ \frac{3}{2}\right)}_{n+1}}{dr} + \left(2 \omega r - \frac{2 l + 3}{r} + \frac{4 \omega r}{(2\omega r^2 + 2l + 3)} \right) \hat{L}^{\left(l+\frac{3}{2}\right)}_{n+1} \right] \nonumber \\
    \fl \qquad  \quad &\quad&\times \frac{N_{n, l+1}\; r^{l+2}}{(2 \omega r^2 + 2 l + 3)}e^{-\frac{1}{2}\omega r^2}.
\label{ope}
\end{eqnarray}
By using the properties and recursion relations involving $X_1$-Laguerre polynomials
the terms inside the square bracket can be replaced by ${\displaystyle -\frac{2}{r}
\left(\frac{2\omega r^2 + 2 l + 3}{2\omega r^2 + 2 l + 1}\right)\; (n+1)\;\hat{L}^{\left(l+\frac{1}{2}\right)}_{n+2}}$ (see {\ref{appa}} for details).
Substituting this result in (\ref{ope}) and simplifying the resultant expression we arrive at
\begin{eqnarray}
\qquad \quad \hat{A}^{+}\Phi_{n,l+1}^{(1)}(r) &=& -\sqrt{4\omega (n+1)}\;\Phi_{n+1,l}^{(1)}(r). \label{susy2b}
\end{eqnarray}

Now let us evaluate the action of other operator $\hat{A}$ on the eigenfunctions $\Phi^{(1)}_{n, l}$, that is
\begin{eqnarray}
\qquad \qquad \quad \hat{A}^{-} \Phi^{(1)}_{n, l} &=& \left(\frac{d}{dr} + W(r)\right)
                          \frac{N_{n, l}\;r^{l+1} \hat{L}^{\left(l+\frac{1}{2}\right)}_{n+1}}{(2\omega r^2 + 2 l + 1)}  e^{-\frac{1}{2} \omega r^2}.  \label{am}
\end{eqnarray}
Here also while expanding the right hand side of (\ref{am}) we find
\begin{eqnarray}
 \fl \hspace{2cm} \hat{A}^{-} \Phi^{(1)}_{n, l}  &=& \left[\frac{d \hat{L}^{\left(l+ \frac{1}{2}\right)}_{n+1}}{dr} - \frac{4 \omega r}{(2\omega r^2 + 2l + 3)} \hat{L}^{\left(l+\frac{1}{2}\right)}_{n+1}\right] \frac{N_{n, l}\; r^{l+1}}{(2 \omega r^2 + 2 l + 1)}  e^{-\frac{1}{2} \omega r^2}.
\label{ope2}
\end{eqnarray}
The terms inside the square bracket can be replaced by
$-2\omega r\left(\frac{2\omega r^2 + 2 l + 1}{2\omega r^2 + 2 l + 3}\right)\;\hat{L}^{\left(l+\frac{3}{2}\right)}_{n}$
(see \ref{appa} again for the details).  With this identification, equation (\ref{am}) can be simplified to
\begin{eqnarray}
  \hat{A}^{-}\Phi_{n,l}^{(1)}(r) &=& -\sqrt{4\omega n}\;\Phi_{n-1,l+1}^{(1)}(r). \label{susy2d}
\end{eqnarray}

Equations (\ref{susy2b}) and (\ref{susy2d}) are
the intertwining relations that relate the system (\ref{hama}) and (\ref{ham1}).
These intertwining operators while commuting yield $[\hat{A}^{-}, \hat{A}^{+}] = 2 W^{'}(r)$. We
recall here that in the case of  harmonic oscillator one finds that
$ W'(r) = 1$ and so the ladder operators associated with the Hamiltonian straightforwardly provide
the Heisenberg-Weyl algebra. As a consequence the ladder operators of the
harmonic oscillator perfectly act as annihilation and creation operators. However, in general,
the factorization operators are not the ladder operators of the system for other than harmonic oscillator.
The latter situation leads to the (nonlinear) polynomial algebras \cite{polyn}. In particular, the
supersymmetric partners of the harmonic oscillator admit distorted versions
of the Heisenberg algebra \cite{hha}. These nonlinear algebras can be linearized through the methods
given in \cite{linear}. However, to the authors knowledge goes, for the shape invariant potentials
other than harmonic oscillator, Balantekin algebraic
method  \cite{bal, bal1} is more versatile to derive the necessary annihilation and creation operators.
These operators frequently lead to either one of the following algebras:
Heisenberg-Weyl algebra, SU(1, 1), SO(2) and $q$-deformed algebra depending on the energy spectrum of the
underlying shape invariant potential. As far as the present problem is concerned, on the Fock space, the commutation
between the operators gives $[\hat{A}^{-}, \hat{A}^{+}] \Phi_{n, l} =  4 \omega \Phi_{n, l}$, which tells us
that  the SUSY operators ${\hat{A}^{\pm}}$ change both the integers $n$ and $l$ (vide equations
(\ref{susy2b}) and (\ref{susy2d})).
To capture perfect raising and lowering operators of the potential ${V_{1}(r)}$
(which should increase and decrease only the number of particles $(n)$ by one
and not the  angular momentum $(l)$ values)
we proceed through the following way.

\subsection{Shape invariance property}
We attempt to  explore the ladder operators by adopting Balentenkin's method \cite{bal}. In this approach,
one can derive the ladder operators by analyzing the shape invariance property
exhibited by the potential under investigation. The potential (\ref{v1}) can be shown to be shape invariant under the condition
\begin{eqnarray}
\qquad \qquad    V_{2}(r;l)=V_{1}(r;l+1)+4\omega.
\end{eqnarray}

From this relation,  we can define two new operators, namely $\hat{B}_{+}$ and $\hat{B}_{-}$,  which are
of the form
\begin{eqnarray}
\fl \qquad \qquad \hat{B}_{+} = \hat{A}^{+} \hat{T}(l) = \hat{A}^{+} e^{\frac{\partial}{\partial l}}, \quad
\qquad \hat{B}_{-} =  \hat{T}^{-1}(l) \hat{A}^{-}  = e^{-\frac{\partial}{\partial l}} \hat{A}^{-}
\end{eqnarray}
respectively.  The commutation relation between these two new operators yields
\begin{eqnarray}
[\hat{B}_{-}, \hat{B}_{+}] = \hat{B}_{-}\hat{B}_{+} - \hat{B}_{+}\hat{B}_{-} = e^{-\frac{\partial}{\partial l}} \hat{A}^{-}\hat{A}^{+} e^{\frac{\partial}{\partial l}} - \hat{A}^{\dagger} \hat{A}.
\label{comm}
\end{eqnarray}

Recalling the operators identity \cite{bch},
\begin{eqnarray}
e^{A} B e^{-A} = B + [A, B] + \frac{1}{2!}[A, [A, B]]+...,
\end{eqnarray}
we can evaluate the first term appearing on the right side in equation (\ref{comm}).
The result shows
\begin{eqnarray}
\fl \quad \qquad e^{-\frac{\partial}{\partial l}} \hat{A}^{-}\hat{A}^{+} e^{\frac{\partial}{\partial l}}
 = -\frac{d^2}{dr^2} + \omega^2 r^2 + \frac{l(l+1)}{r^2} + \frac{8\omega }{(2\omega r^2 + 2l+1)} \nonumber \\
\qquad \qquad \qquad - \frac{16 \omega (2l+1)}{(2\omega r^2 + 2l+ 1)} - \omega (2 l  - 1).
\label{res}
\end{eqnarray}
On the otherhand evaluating $\hat{A}^{+}\hat{A}^{-}$  we find
\begin{eqnarray}
\fl \;\hat{A}^{+}\hat{A}^{-}= -\frac{d^2}{dr^2} + \omega^2 r^2 + \frac{l(l+1)}{r^2} + \frac{8\omega }{(2\omega r^2 + 2l+1)} - \frac{16 \omega (2l+1)}{(2\omega r^2 + 2l+ 1)} - \omega (2 l+3).
\label{res2}
\end{eqnarray}
Subtracting (\ref{res2}) from (\ref{res}), we get
\begin{eqnarray}
\qquad \qquad \qquad [\hat{B}_{-}, \hat{B}_{+}] =  4\omega.
\label{alg}
\end{eqnarray}

The relation (\ref{alg}) confirms that the operators $\hat{B}_{+}$ and
$\hat{B}_{-}$ form Heisenberg-Weyl algebra, which may also be observed from the
energy spectrum of the potential $V_1$, given in (\ref{en1}),
which is linear (vide equation (\ref{en1})). We note here that the
new operators $\hat{B}_{+}$ and $\hat{B}_{-}$ act
as the ladder operators of the extended
radial oscillator potential as in the case of harmonic oscillator potential  \cite{bal}.

From the above we can also establish
\begin{eqnarray}
  \hat{B}_{-} \Phi_{n,l}^{(1)}= -\sqrt{4\omega n}\;\Phi_{n-1,l}^{(1)}, \quad \quad
  \hat{B}^{+} \Phi_{n,l}^{(1)}= -\sqrt{4\omega(n+1)}\;\Phi_{n+1,l}^{(1)}.
\end{eqnarray}
Further, redefining these two operators,  $\hat{B}_{-}$ and $\hat{B}_{+}$, in such a way that
\begin{eqnarray}
  \hat{a}_{r} = -\frac{1}{\sqrt{4\omega}}\;\hat{B}_{-}, \quad \quad
   \hat{a}^{\dagger}_{r} = -\frac{1}{\sqrt{4\omega}}\;\hat{B}_{+},
\end{eqnarray}
we can show that
\begin{eqnarray}
  \hat{a}_{r} \Phi_{n,l}^{(1)}=\sqrt{n}\;\Phi_{n-1,l}^{(1)}, \quad \quad
  \hat{a}^{\dagger}_{r} \Phi_{n,l}^{(1)}=\sqrt{n+1}\;\Phi_{n+1,l}^{(1)}.
\label{ancr}
\end{eqnarray}
The above relations confirm that these operators perfectly annihilate and create the eigenstate  $\Phi_{n,l}^{(1)}$ by absorbing and
emitting one photon.

Using these ladder operators,  one can construct coherent, squeezed and other type of states
and analyze classical/non-classical properties
exhibited by the radial oscillator potential (\ref{poten}). In the following we
carry out these studies.

\section{Squeezed coherent states}
In this section, using the above ladder operators, we construct squeezed coherent states
of the three dimensional isotonic oscillator.  The squeezed coherent states of the three dimensional spherically
symmetric oscillator can be obtained by taking a tensor product of squeezed coherent states of radial
excitation with squeezed angular momentum coherent states \cite{nito}, that is
\begin{eqnarray}
\fl\hspace{5cm}|\xi,\alpha \rangle &=& |\xi_r,\alpha_{r}  \rangle \otimes | \tilde\xi, \tilde\alpha \rangle.
\label{scs1}
\end{eqnarray}

The squeezed coherent states which generalize both the coherent and squeezed states are defined to be  \cite{scs}
\begin{eqnarray}
|z, \alpha\rangle = \hat{S}(z)\;\hat{D}(\alpha) |0\rangle = \hat{D}(\alpha_0)\;\hat{S}(z) |0\rangle,
\label{scsop}
\end{eqnarray}
where $\hat{S}(z) = \exp{\left[\frac{1}{2}(z^{*} \hat{a}^{2} - z \hat{a}^{{\dagger}^2})\right]}$ is the squeezing operator with
$z = R e^{i \phi}$ and $\hat{D}(\alpha) = \exp{(\alpha {\hat{a}^{\dagger}} - \alpha^{*} \hat{a})}$ in which the operators
$\hat{a}$ and $\hat{a}^{\dagger}$ represent annihilation and creation operators respectively. These
operators satisfy
the relations given in (\ref{ancr}).
Here $\alpha =  \alpha_0\cosh{R}  +  \alpha^{*}_{0}\;e^{i \phi}  \sinh{R}$,
where $R$ and $\phi$ are the squeezing parameters and $\alpha_0$ and $\alpha^{*}_0$ are
the coherent parameters respectively.

\subsection{Radial part}
To begin with we evaluate the squeezed coherent states of
the radial part, that is $|\xi_r, \alpha_r\rangle$,
by using the
definition
\begin{eqnarray}
|\xi_r, \alpha_r \rangle = \hat{D}(\alpha_{0, r}) \hat{S}(z_r)\;|0\rangle.
\label{scs_r}
\end{eqnarray}
The squeezing operator can be disentangled as  \cite{scs}
\begin{eqnarray}
\hat{S}(z_r) = \frac{1}{\cosh{R_r}}e^{-\frac{1}{2}\tanh{R_r}\;e^{i \phi_r} \hat{a}_{r}^{{\dagger}^2}} e^{-\ln\cosh{R_r}\;\hat{a}_{r}^{\dagger} \hat{a}_r }
e^{\frac{1}{2}\tanh{R_r}\;e^{ -i \phi_r} \hat{a}_{r}^2}.
\label{dis1a}
\end{eqnarray}
From (\ref{dis1a}), we find
\begin{eqnarray}
\hat{S}(z_r)|0\rangle = \frac{1}{\cosh{R_r}}e^{-\frac{1}{2}e^{i \phi_r}\;\tanh{R_r}\; \hat{a}_{r}^{{\dagger}^2}} |0\rangle.
\label{dis1}
\end{eqnarray}
Substituting (\ref{dis1}) in (\ref{scs_r}), we obtain
\begin{eqnarray}
|\xi_r, \alpha_r\rangle = \frac{1}{\cosh{R_r}} \hat{D}(\alpha_{0,r})\;e^{\frac{\xi_r}{2} \hat{a}_{r}^{\dagger^2}} |0\rangle,
\label{scsop1}
\end{eqnarray}
where  we have defined $\xi_r = -e^{ i \phi_r}\;\tanh{R_r}$. Using the identity
 $\hat{D}^{\dagger}(\alpha_{0,r}) \hat{D}(\alpha_{0, r}) = 1$, we can
rewrite (\ref{scsop1}) of the form
\begin{eqnarray}
|\xi_r, \alpha_r\rangle = \frac{1}{\cosh{R_r}} \hat{D}(\alpha_{0,r}) e^{\frac{\xi_r}{2} \hat{a}_{r}^{\dagger^2}} \hat{D}^{\dagger}(\alpha_{0,r})
                            \hat{D}(\alpha_{0,r}) |0\rangle. \label{opd}
\end{eqnarray}
The combined action of $\hat{D}(\alpha_{0, r})\;e^{\frac{\xi_r}{2} \hat{a}_{r}^{{\dagger}^2}}\;\hat{D}^{\dagger}(\alpha_{0, r})$,
yields
\begin{eqnarray}
\fl \qquad \qquad \qquad \hat{D}(\alpha_{0, r}) e^{\frac{\xi_r}{2} \hat{a}_{r}^{{\dagger}^2}} \hat{D}^{\dagger}(\alpha_{0, r}) = e^{\frac{\xi_r}{2} (\hat{a}_{r}^{\dagger}-\alpha^*_{0, r})^2} \approx e^{\frac{\xi_r}{2} \hat{a}_{r}^{\dagger^2}-\alpha^*_{0, r} \xi_r \hat{a}_{r}^{\dagger} }.
\label{c1}
\end{eqnarray}
Substituting this result in (\ref{opd}) and simplifying the latter we obtain
\begin{eqnarray}
\fl \qquad \quad |\xi_r, \alpha_r\rangle = \frac{1}{\cosh{R_r}} e^{\left(\frac{\xi_r}{2} \hat{a}_{r}^{\dagger^2} + (\alpha_{0, r} - \alpha^{*}_{0, r} \xi_r) \hat{a}^{\dagger}_r \right)}|0\rangle  = \frac{1}{\cosh{R_r}} e^{\left(\frac{\xi_r}{2} \hat{a}_{r}^{\dagger^2} + \frac{\alpha_{r}}{\cosh{R_r}} \hat{a}^{\dagger}_r \right)} |0\rangle,
\label{opsq}
\end{eqnarray}
where  $\alpha_{r} =  \cosh{R_r}\; \alpha_{0, r} + e^{i\phi_r} \sinh{R_r}\; \alpha^{*}_{0, r}$.

We can replace the exponential part in (\ref{opsq}) by Hermite functions,
$H_{n}$, $n = 0, 1, 2, 3, ...$, that is \cite{book}
\begin{eqnarray}
e^{\left(\frac{\xi_r}{2} \hat{a}_{r}^{\dagger^2} + \frac{\alpha_{r}}{\cosh{R_r}} {\hat{a}}^{\dagger}_r\right)}
= \sum^{\infty}_{n = 0}\frac{H_{n}\left(\frac{\alpha_r \sqrt{1 - |\xi_r|^2}}{\sqrt{-2 \xi_r}}\right)}{n!} \left(-\frac{\xi_r}{2}\right)^{n/2} \hat{a}^{{\dagger}^n}_r
\label{her}
\end{eqnarray}
so that equation (\ref{opsq}) now becomes
\begin{eqnarray}
\fl\qquad \qquad \qquad |\xi_{r},\alpha_{r}\rangle &=& \sum^{\infty}_{n = 0}\frac{H_{n}\left(\frac{\alpha_r \sqrt{1 - |\xi_r|^2}}{\sqrt{-2 \xi_r}}\right)}{n!} \left(-\frac{\xi_r}{2}\right)^{n/2} \hat{a}^{{\dagger}^n}|0\rangle, \nonumber \\
                               &=& \sum_{n=0}^{\infty}\frac{H_{n}\left(\frac{\alpha_{r}\sqrt{1 - |\xi_r|^2}}{\sqrt{-2\xi_{r}}}\right)}{\sqrt{n!}}\left(-\frac{\xi_r}{2}\right)^{n/2}| n\rangle.
\label{scsr}
\end{eqnarray}

Thus the normalized squeezed coherent states of the radial part are found to be
\begin{eqnarray}
|\xi_{r},\alpha_{r}\rangle  =  N_{r}\sum_{n=0}^{\infty}\frac{H_{n}\left(\frac{\alpha_{r}\sqrt{1 - |\xi_r|^2}}{\sqrt{-2\xi_{r}}}\right)}{\sqrt{n!}}\left(-\frac{\xi_r}{2}\right)^{n/2}| n\rangle,
\label{scsr}
\end{eqnarray}
where $N_{r}$ is the normalization constant whose exact value is given by
\begin{eqnarray}
N_{r} = (1 - |\xi_r|^2)^{1/4} \exp{\left[-\frac{1}{4}(\alpha^2_r \xi^{*}_r + \alpha^{{*}^2}_{r} \xi_{r} + 2 |\alpha_r|^2)\right]}, \quad  0 < |\xi_r| < 1.
\label{nr}
\end{eqnarray}

In the following, we construct squeezed coherent states of the angular part.

\subsection{Angular momentum part}
The angular momentum part of the Schr\"{o}dinger equation (\ref{ddim}) with
$d = 3$ admits spherical harmonics $Y_{l, m}(\theta, \phi)$ as the solution which
is defined to be simultaneous eigenstates of the operators $\hat{L}^2$ and $\hat{L}_z$
with the eigenvalues $l(l+1)$ and $m$, that is
\begin{eqnarray}
    \hat{L}^2| l,m\rangle &=& l(l+1) | l,m\rangle,\\
    \hat{L}_z| l,m\rangle &=& m| l,m\rangle,
\end{eqnarray}
where  $\langle \theta ,\phi|l,m \rangle =Y_{l,m}(\theta ,\phi )$ and $|l, m\rangle$ are nothing but the
angular momentum states.

The angular momentum coherent states can be constructed by expressing the
angular momentum coordinates, $\hat{L}_{+}, \hat{L}_{-}$, $\hat{L}_{z}$ and
$\hat{L}^2$, in terms of the new annihilation and creation operators, that is
$\hat{a}_{\pm}$ and $\hat{a}^{\dagger}_{\pm}$ \cite{schwinger, sch}:
\begin{eqnarray}
    \hat{L}_{\pm} &=& \hat{a}_{\pm}^{\dagger} \hat{a}_{\mp }, \label{lpp}\\
    \hat{L}_{z}  &=& \frac{1}{2} \left(\hat{a}^{\dagger}_{+} \hat{a}_{+} - \hat{a}^{\dagger}_{-} \hat{a}_{-}\right)
             = \frac{1}{2} (\hat{n}_{+} - \hat{n}_{-}), \label{lzz}\\
     \hat{L}^2    &=& \frac{1}{2} \left(\hat{n}_{+} + \hat{n}_{-}\right)\left(\frac{1}{2} \left(\hat{n}_{+} + \hat{n}_{-}\right)+1\right),
\end{eqnarray}
where $ \hat{n}_{+} + \hat{n}_{-} = \hat{n} $ is the number operator. The action of the operators $\hat{a}_{+}$
and $\hat{a}_{-}$ on the number states is given by \cite{schwinger}
\begin{eqnarray}
 \hat{a}_{+}\; |n\rangle  &=& \sqrt{n_{+}}\;|n_{+}-1,n_{-}\rangle, \qquad \;\;\; \hat{a}_{-}\;|n\rangle  = \sqrt{n_{-}}|  n_{+},n_{-}-1\rangle,
 \label{ops}
\\
 \hat{a}_{+}^{\dagger}\; | n\rangle &=& \sqrt{n_{+}+1}\;| n_{+}+1, n_{-}\rangle, \quad \hat{a}_{-}^{\dagger}\; | n\rangle = \sqrt{n_{-}+1}\;| n_{+},
n_{-}+1\rangle.
\label{op}
\end{eqnarray}
The operator $\hat{L}_{-}$ act on the states ${|l,m\rangle}$ as follows: 
\begin{eqnarray}
    \hat{L}_{-}| l,m\rangle =\sqrt{(l+m)(l-m+1)}\;| l,m-1\rangle.
\end{eqnarray}

The squeezed  coherent states associated with the angular momentum part can be  obtained by using the definition
\cite{abir},
\begin{eqnarray}
  |\tilde\xi ,\tilde\alpha \rangle &=& \hat{D}(\tilde\alpha_0)\hat{S}(\tilde\xi) | 0,0 \rangle.
\end{eqnarray}
Since we are dealing two modes, we consider the above equation is of the form
\begin{eqnarray}
  | \tilde\xi ,\tilde\alpha \rangle & =& \hat{D}_{+}(\alpha_{0,+})\hat{S}_{+}( \xi_{+})\hat{D}_{-}( \alpha_{0,-})
\hat{S}_{-}( \xi_{-}) | 0,0 \rangle,
  \label{s0}
\end{eqnarray}
with $\xi_{\pm} =  -\tanh{R_{\pm}}\;e^{ i \phi_{\pm}}$  and
$\alpha_{\pm} = \alpha_{0,\pm} \cosh{R_{\pm}}\; + \alpha^{*}_{0, \pm} \sinh{R_{\pm}}\; e^{i\phi_{\pm}} $
are the squeezing parameters and displaced coherent parameters respectively.

By adopting the calculations given in the radial part we find the combined action of
displacement and squeezed operators produce (vide equations (\ref{ops}) and (\ref{op}))
\begin{eqnarray}
  \hat{D}_{+}(\alpha_{0,+})\hat{S}_{+}(\xi_{+})=\exp{\left(\frac{\alpha_{+}}{\cosh{R_{+}}}\hat{a}_{+}^{\dagger}+\frac{\xi_{+}}{2}
\hat{a}_{+}^{{\dagger}^2}\right)}.
\label{dis}
\end{eqnarray}
Recalling the identity, ${\displaystyle \sum^{\infty}_{n = 0} H_{n}(x) \frac{t^{n}}{n!} = e^{2tx - t^2}}$,
we can express the operator appearing in (\ref{dis})  as  an infinite series in Hermite polynomials, that is
\begin{eqnarray}
  \hat{D}_{+}(\alpha_{0,+})\hat{S}_{+}(\xi_{+})=\sum _{n_+=0}^{\infty}\frac{H_{n_+}\left(\frac{\alpha_{+}\sqrt{1 - |\xi_{+}|^2}}{\sqrt{-2\xi_{+}}}\right)}{{n_+}!}\left(-\frac{\xi_{+}}{2}\right)^{{n_+}/2}\hat{a}_{+}^{{\dagger}^{n_{+}}}.
  \label{s1}
\end{eqnarray}
By induction, we find
\begin{eqnarray}
  \hat{D}_{-}(\alpha_{0,-})\hat{S}_{-}(\xi_{-})=\sum _{n_{-}=0}^{\infty}\frac{H_{n_-}\left(\frac{\alpha_{-}\sqrt{1 - |\xi_{-}|^2}}{\sqrt{-2\xi_{-}}}\right)}{n_{-}!}\left(-\frac{\xi_{-}}{2}\right)^{n_{-}/2}\hat{a}_{-}^{{\dagger}^{n_{-}}}.
  \label{s2}
\end{eqnarray}
Substituting (\ref{s1}) and (\ref{s2}) in (\ref{s0}), we obtain
\begin{eqnarray}
\fl \qquad |\tilde\xi,\tilde\alpha \rangle = N_{\pm} \sum_{n_{+},n_{-}=0}^{\infty}\frac{H_{n_{+}}\left(\frac{\alpha_{+}\sqrt{1 - |\xi_{+}|^2}}{\sqrt{2\xi_{+}}}\right)H_{n_{-}}\left(\frac{\alpha_{-}\sqrt{1 - |\xi_{-}|^2}}{\sqrt{-2\xi_{-}}}\right)}{n_{+}! n_{-}!} \nonumber \\
\qquad \qquad \qquad \times{\left(\frac{-\xi_{+}}{2}\right)}^{n_{+}/2}{\left(\frac{-\xi_{+}}{2}\right)}^{n_{-}/2}
\hat{a}_{+}^{{\dagger}^{n_{+}}}\hat{a}_{-}^{{\dagger}^{n_{-}}}|0,0\rangle, \label{ascs}  \;\;
\end{eqnarray}
where  $N_{\pm}$ is the normalization constant which can be fixed as
\begin{eqnarray}
\hspace{-0.5cm} N_{\pm} = (1 - |\xi_+|^2)^{1/4}\;(1 - |\xi_-|^2)^{1/4} \exp{\left[-\frac{1}{4}(\alpha^2_{+} \xi^{*}_{+} + \alpha^{{*}^2}_{+} \xi_{+} + 2 |\alpha_+|^2) \right]} \nonumber \\
\qquad \qquad \;\; \times \exp{\left[-\frac{1}{4}(\alpha^2_{-} \xi^{*}_{-} + \alpha^{{*}^2}_{-} \xi_{-} + 2 |\alpha_{-}|^2)\right]}, \;
0 < |\xi_{\pm}| < 1,
\label{npm}
\end{eqnarray}
through the usual procedure.

Using the identity,
\begin{eqnarray}
 \qquad \quad  | l,m\rangle=\frac{(\hat{a}_{+}^{\dagger})^{l+m}(\hat{a}_{-}^{\dagger})^{l-m}}{\sqrt{(l+m)!(l-m)!}} | 0,0 \rangle,
\label{lm}
\end{eqnarray}
with the restriction $n_{+} = l + m$ and $n_{-} = l - m$,
we can rewrite the squeezed angular momentum coherent states  (\ref{ascs}) in the form
\begin{eqnarray}
\fl\quad \qquad \quad  |\tilde \xi, \tilde\alpha \rangle = N_{\pm}\sum_{l=0}^{\infty} \sum_{m=-l}^{l} \frac{H_{l+m}\left(\frac{\alpha_{+}\sqrt{1 - |\xi_{+}|^2}}{\sqrt{-2\xi_{+}}}\right)H_{l+m}\left(\frac{\alpha_{-}\sqrt{1 - |\xi_{-}|^2}}{\sqrt{-2\xi_{-}}}\right)}{\sqrt{(l+m)!(l-m)!}} \nonumber \\
\qquad \qquad \qquad \quad \times\left(\frac{-\xi_{+}}{2}\right)^{\frac{l+m}{2}} \left(\frac{-\xi_{-}}{2}\right)^{\frac{l-m}{2}} | l,m\rangle.
\label{scsa}
\end{eqnarray}

Finally, substituting  the squeezed coherent states of
the radial part (vide equation (\ref{scsr})) and the angular part (vide equation
(\ref{scsa})) in (\ref{scs1}) we obtain the
squeezed coherent states of the three dimensional system (\ref{3_dim})
which in turn reads
\begin{eqnarray}
\fl\;\;| \xi, \alpha \rangle = N_{\xi, \alpha}\sum_{n=0}^{\infty}\sum_{l=0}^{\infty}\sum_{m=-l}^{l}\frac{H_{n}\left(\frac{\alpha_r \; \sqrt{1 - |\xi_r|^2}}{\sqrt{-2 \xi_r}}\right)H_{l+m}\left(\frac{\alpha_{+} \; \sqrt{1 - |\xi_{+}|^2}}{\sqrt{-2 \xi_{+}}}\right)H_{l-m}\left(\frac{\alpha_{-} \; \sqrt{1 - |\xi_{-}|^2}}{\sqrt{-2 \xi_{-}}}\right)}{\sqrt{n!(l+m)!(l-m)!}}\nonumber \\
\quad \times \left(\frac{-\xi_{r}}{2}\right)^{\frac{n}{2}}\left(\frac{-\xi_{+}}{2}\right)^{\frac{l+m}{2}} \left(\frac{-\xi_{-}}{2}\right)^{\frac{l-m}{2}}|n, l,m \rangle,\; \nonumber \\
 \qquad \quad 0 < |\xi_{r}| < 1, \quad 0 < |\xi_{\pm}| < 1.
\label{scs}
\end{eqnarray}
The normalization constant, $N_{\xi, \alpha}$, is given by
\begin{eqnarray}
\fl \;\;\;  N_{\xi, \alpha} = N_{r} \times  N_{\pm} \nonumber \\
\fl \qquad \quad               =\left[(1 - |\xi_r|^2)\;(1 - |\xi_+|^2)\;(1 - |\xi_-|^2)\right]^{1/4} \exp{\left[-\frac{1}{4}(\alpha^2_{r} \xi^{*}_{r} + \alpha^{{*}^2}_{r} \xi_{r} + 2 |\alpha_r|^2) \right]} \nonumber \\
              \hspace{-0.6cm} \times  \exp{\left[-\frac{1}{4}(\alpha^2_{+} \xi^{*}_{+} + \alpha^{{*}^2}_{+} \xi_{+} + 2 |\alpha_+|^2) \right]} \exp{\left[-\frac{1}{4}(\alpha^2_{-} \xi^{*}_{-} + \alpha^{{*}^2}_{-} \xi_{-} + 2 |\alpha_{-}|^2)\right]}, \; \nonumber \\
\qquad  0 < |\xi_r| < 1, \;\;0 < |\xi_{\pm}| < 1.
\end{eqnarray}
Equation (\ref{scs}) can also be written in a more compact form, namely
\begin{eqnarray}
|\xi,\alpha \rangle= N_{\xi, \alpha}\sum_{n=0}^{\infty}\sum_{l=0}^{\infty}\sum_{m=-l}^{l} c_{n} c_{l,m} |n, l, m\rangle,
\label{scsf}
\end{eqnarray}
with
\begin{subequations}
\begin{eqnarray}
\fl \qquad c_{n} =  \frac{H_{n}\left(\frac{\alpha_r \; \sqrt{1 - |\xi_r|^2}}{\sqrt{-2 \xi_r}}\right)}{\sqrt{n!}}\left(\frac{-\xi_{r}}{2}\right)^{\frac{n}{2}} , \label{cn}
\end{eqnarray}
\mbox{and}
\begin{eqnarray}
\fl \qquad c_{l, m} =  \frac{ H_{l+m}\left(\frac{\alpha_{+} \; \sqrt{1 - |\xi_{+}|^2}}{\sqrt{-2 \xi_{+}}}\right) H_{l+m}\left(\frac{\alpha_{-} \; \sqrt{1 - |\xi_{-}|^2}}{\sqrt{-2 \xi_{-}}}\right)}{\sqrt{(l+m)!(l-m)!}}\left(-\frac{\xi_{+}}{2} \right)^{\frac{l+m}{2}} \left(-\frac{\xi_{-}}{2}\right)^{\frac{l-m}{2}}.\hspace{1cm} \label{clm}
\end{eqnarray}
\end{subequations}

The squeezed coherent states given in (\ref{scsf}) are expressed  in terms of both number states ($|n\rangle$)
and angular momentum states $(|l, m\rangle)$, that is  the states are
expressed in terms of bound state solutions $\psi(r, \theta, \phi)$. In other words a
complete description of the three mode squeezed coherent states is established now in terms of
three independent squeezing parameters $(\xi_r, \xi_+, \xi_-)$ and
three independent displaced coherent  parameters $(\alpha_r, \alpha_+, \alpha_-)$.
By varying these parameters we can analyze both the
classical and non-classical nature of these states.
We recall here that whenever the quantum states exhibit Poissonian statistics 
they said to possess classical nature \cite{how}. Any deviation from this 
behaviour represents the non-classical nature of the states.
To proceed further  we restrict the range of values of
$\xi_j$ to be $0< |\xi_j| < 1$ and the range ${\alpha_{0,j}}^{'}$s be $-\infty$ to $\infty$
since normalization constant $N_{\xi, \alpha}$ is defined in the
range $0 < |\xi_j| < 1$, where $j = r, +, -$.

In the following section, we analyze the classical/non-classical nature of the obtained squeezed coherent states.

\subsection{\bf Completeness condition}
In this sub-section, we demonstrate that the three
mode squeezed coherent states (\ref{scsf}) satisfy the completeness condition.
The three mode squeezed coherent states  are represented in terms of three
coherent parameters (${\alpha^{'}_{0,j}} s , \; j = r, +, -$)
and three squeezed parameters $(\xi_{j}'s, \; j = r, +, -)$. Now
we prove that the squeezed coherent states resolve the identity operator \cite{rii}
\begin{eqnarray}
\left(\frac{i}{2 \pi}\right)^3 \int |\xi, \alpha\rangle\langle \xi, \alpha|\; d^{2} \alpha = \hat{I},
\label{ri1}
\end{eqnarray}
when the integration is carried over the entire space of $\alpha$ $(\alpha_{r}, \alpha_{+}, \alpha_{-})$. In this
analysis we consider $\alpha$  varies only with respect to $\alpha^{'}_{0, j}$s and treat $\xi^{'}_j$s
are arbitrary constants.

For the squeezed coherent parameters,
\begin{eqnarray}
\hspace{1cm} \alpha_j = \frac{\alpha_{0, j} - \xi_j \alpha^{*}_{0,j}}{\sqrt{1 - |\xi_j|^2}}, \quad j = r, \theta, \phi,
\label{ri1a}
\end{eqnarray}
we can show that
\begin{eqnarray}
d\alpha_j d\alpha^{*}_j =  \left|\begin{array}{cc}
    \frac{\partial \alpha_j}{\partial \alpha_{0,_j}} &  \frac{\partial \alpha_j}{\partial \alpha^*_{0,j}} \\
    \frac{\partial \alpha^{*}_{j}}{\partial \alpha_{0,j}} & \frac{\partial \alpha^{*}_{j}}{\partial \alpha^{*}_{0,j}}  \\
        \end{array} \right| d\alpha_{0,j} d\alpha^{*}_{0,j} = d\alpha_{0,j} d\alpha^{*}_{0,j}.
\label{ri2}
\end{eqnarray}

Substituting the expression given in (\ref{scsf}) in  the integral (\ref{ri1}),
we get
\begin{eqnarray}
\fl \hspace{0.5cm}G = \int |\xi, \alpha\rangle\langle \xi, \alpha|\; d^{2} \alpha = \sum^{\infty}_{n'=0}\sum^{\infty}_{l'=0} \sum^{l'}_{m'=-l'}\sum^{\infty}_{n=0}\sum^{\infty}_{l=0} \sum^{l}_{m=-l}
|n', l', m'\rangle \langle n, l, m|  \nonumber \\
\hspace{3cm} \times \int \int \int c^*_{n'} c^*_{l', m'}  c_{n}  c_{l,m} N^{2}_{\xi, \alpha} d^{2}\alpha_r d^2\alpha_{+} d^{2}\alpha_{-}.
\label{ri3}
\end{eqnarray}

To start with we evaluate the radial part (which we call as $G_R$) in (\ref{ri3}), that is
\begin{eqnarray}
\fl \quad G_R &=& \sum^{\infty}_{n = 0}\sum^{\infty}_{n' = 0} |n'\rangle \langle n| \int c^{*}_{n'} c_{n} N^{2}_{r} d^{2} \alpha_r \nonumber \\
\fl  &=& \sum^{\infty}_{n = 0}\sum^{\infty}_{n' = 0} \frac{|n'\rangle\langle n|}{\sqrt{n'!\;n!}}\sqrt{1 - |\xi_r|^2}
\left(-\frac{\xi^{*}_{r}}{2}\right)^{n'/2} \left(-\frac{\xi_{r}}{2}\right)^{n/2}\int \exp{\left[-|\alpha_r|^2 -\frac{1}{2}\left(
\xi_r \alpha^{{*}^2}_{r}+\xi^{*}_r \alpha^{2}_{r}\right)\right]}\nonumber \\
\fl & &\times  H_{n'}\left(\frac{\alpha^{*}_r \; \sqrt{1 - |\xi_r|^2}}{\sqrt{-2 \xi^*_r}}\right)H_{n}\left(\frac{\alpha_r \; \sqrt{1 - |\xi_r|^2}}{\sqrt{-2 \xi_r}}\right)d\alpha_r d\alpha^{*}_r.
\label{ri4}
\end{eqnarray}
Substituting (\ref{ri1a}) and (\ref{ri2}) in (\ref{ri4}) and separating the integral
(which we call as $I_1$) from the summation part, we get
\begin{eqnarray}
\fl \qquad \;   I_1 &=& \int \int \exp{\left[-|\alpha_{o,r}|^2 + \frac{1}{2}\left(\xi_r \alpha^{*^2}_{0,r} + \xi^{*}_{r} \alpha^2_{0,r} \right)\right]}\nonumber \\
\fl \qquad \;  & &  \times H_{n'}\left(\frac{\alpha^{*}_{0,r}}{\sqrt{-2\xi^*_r}} +\sqrt{-\frac{\xi^*_r}{2}} \alpha_{0,r}\right)
H_{n}\left(\frac{\alpha_{0,r}}{\sqrt{-2\xi_r}} +\sqrt{-\frac{\xi_r}{2}} \alpha^*_{0,r}\right)d\alpha_{0,r} d\alpha^{*}_{0,r}.
\label{r5}
\end{eqnarray}

To evaluate the integral (\ref{r5}) we introduce the transformation
\begin{eqnarray}
\frac{\alpha_{0, j}}{\sqrt{- 2 \xi_j}} = \frac{y_{j} + |\xi_j| y^{*}_{j}}{1-|\xi_j|^2}\quad\mbox{and}\quad
\frac{\alpha^{*}_{0, j}}{\sqrt{- 2 \xi_j}} = \frac{y^{*}_{j} + |\xi_i| y_{j}}{1-|\xi_j|^2}
\label{r5a}
\end{eqnarray}
so that
\begin{eqnarray}
\fl \qquad d\alpha_{0,j} d\alpha^{*}_{0,j} =   \left|\begin{array}{cc}
    \frac{\partial \alpha_{0,j}}{\partial y_j} &  \frac{\partial \alpha_{0,j}}{\partial y^*_j} \\
    \frac{\partial \alpha^{*}_{0,j}}{\partial y_j} & \frac{\partial \alpha^{*}_{0,j}}{\partial y^{*}_j}  \\
        \end{array} \right|  dy_j dy^{*}_j = \frac{-2|\xi_j|}{1 - |\xi_j|^2} dy_j dy^{*}_j,\quad \; j = r, \theta, \phi.
\label{ri6}
\end{eqnarray}
In the new variables the integral (\ref{r5}) reads
\begin{eqnarray}
\fl \qquad I_1 &=& \frac{-2|\xi_r|}{1-|\xi_r|^2} \int \int \exp{\left[\frac{|\xi_r|^2 (y^2_r + y^{*^2}_r) + 2 |\xi_r| |y_r|^2}{1 - |\xi_r|^2}\right]}
H_{n}(y_r)H_{n'}(y^*_r) dy_r dy^*_r.
\label{r7}
\end{eqnarray}

With another change of variable, $z = i\frac{|\xi_r|}{\sqrt{1 - |\xi_r|^2}} y_r$,
the integral (\ref{r7}) can be brought to the form
\begin{eqnarray}
\fl \qquad I_1 =  \frac{-2 i}{\sqrt{1-|\xi_r|^2}} \int \exp{\left[\frac{|\xi_r|^2 y^{*^2}_r}{1 - |\xi_r|^2}\right]} H_{n}[y^*_r] \int \exp{\left[-z^2 + 2 z \frac{-i y^*_r}{\sqrt{1 - |\xi_r|^2} }\right]} \nonumber \\
\hspace{3cm} \times  H_{n}\left[\frac{-i\sqrt{1-|\xi_r|^2}}{|\xi_r|} z\right] dz\;dy^*_r.
\label{r8}
\end{eqnarray}
With the identity  \cite{book}
\begin{eqnarray}
\int e^{-(x-y)^2} H_n[\alpha x] dx = \sqrt{\pi} (1 - \alpha^2)^{n/2} H_{n}\left[\frac{\alpha}{\sqrt{1-\alpha^2}}y\right]
\label{r9}
\end{eqnarray}
the second integral in (\ref{r8}) can be evaluated and given in terms of Hermite polynomials, that is,
\begin{eqnarray}
\fl \qquad \qquad I_1 =  \frac{-2 i}{\sqrt{1-|\xi_r|^2}} \sqrt{\pi} \left(\frac{1}{|\xi_r|}\right)^{n}
\int \exp{\left[\frac{(|\xi_r|^2-1) y^{*^2}_r}{1 - |\xi_r|^2}\right]} H_{n}[-y^*_r] H_{n'}[y^*_r] dy^{*}_r.
\label{r10}
\end{eqnarray}
Now using the orthogonality property of the Hermite polynomials,
\begin{eqnarray}
 \qquad\int e^{-x^2} H_n[x] H_n'[x]dx = \sqrt{\pi}\;2^n\;n!\;\delta_{n,n'},
\label{r11}
\end{eqnarray}
where $\delta_{n,n'}$ is Kronecker-delta function, we find
\begin{eqnarray}
\fl \qquad \qquad \qquad  \qquad I_1 =   \frac{- 2 \pi i}{\sqrt{1-|\xi_r|^2}}  \left(-\frac{2}{|\xi_r|}\right)^{n} n! \delta_{n,n'}.
\label{r12}
\end{eqnarray}
Substituting this expression, (\ref{r12}), in $G_R$ (vide equation  (\ref{ri4})) and simplifying the
resultant expression,  we get
\begin{eqnarray}
\fl \qquad \qquad  G_R &=& - 2 \pi i \sum^{\infty}_{n = 0}\sum^{\infty}_{n' = 0} \frac{|n'\rangle\langle n|}{\sqrt{n'!n!}} \left(-\frac{\xi^{*}_{r}}{2}\right)^{n'/2} \left(-\frac{\xi^{*}_{r}}{2}\right)^{n/2}  \left(-\frac{2}{|\xi_r|}\right)^{n} n! \delta_{n,n'}.
\label{r12a}
\end{eqnarray}
From the above we observe that radial part of the integral provides
\begin{eqnarray}
\fl \qquad \qquad \qquad \qquad G_R = - 2  i \pi \sum^{\infty}_{n = 0} |n\rangle \langle n|.
\label{r13}
\end{eqnarray}

In a similar way we can  evaluate the integrals involving the other two variables, namely $\theta$ and
$\phi$ (since the procedure is exactly the same as that of radial part  we do not repeat the details here). Our result shows that
\begin{eqnarray}
\fl \qquad  G_A &=& \sum^{\infty}_{l'=0} \sum^{l'}_{m'=-l'}\sum^{\infty}_{l=0} \sum^{l}_{m=-l}
|l', m'\rangle \langle l, m|  \int \int \int c^*_{n'} c^*_{l', m'}  c_{n}  c_{l,m} N^{2}_{\pm} d^{2}\alpha_r d^2\alpha_{+} d^{2}\alpha_{-} \nonumber \\
\fl \qquad  \; &=& -4 \pi^2 \sum^{\infty}_{l = 0} \sum^{l}_{m=-l} |l,m\rangle \langle l,m|.
\label{r14}
\end{eqnarray}

The  integral (\ref{ri3}) is completely evaluated now. The resultant value turns out to be $G = G_R \times G_A =
 i 8 \pi^3 \sum^{\infty}_{n = 0} \sum^{\infty}_{l = 0} \sum^{l}_{m=-l}$
$|l,m\rangle |n \rangle \langle n| \langle l,m|$. This in turn confirms
that the states $|\xi,\alpha\rangle$ resolve the identity operator, that is
\begin{eqnarray}
\qquad \qquad \sum^{\infty}_{n = 0} \sum^{\infty}_{l = 0} \sum^{l}_{m=-l} |n, l, m\rangle \langle n, l, m| = \hat{I}.
\label{r15}
\end{eqnarray}
The result ensures that the constructed three mode squeezed coherent states form a complete set.

\section{\bf Non-classical properties}
\subsection{\bf Mandel's $Q$ parameter}
In this sub-section, we study Mandel's $Q$ parameter for the three mode
squeezed coherent states (\ref{scsf}). We  evaluate
Mandel's $Q$ parameter associated with each mode \cite{Mandel}
\begin{eqnarray}
Q_j = \frac{\langle \hat{n}^2_{j} \rangle}{\langle \hat{n}_{j} \rangle} - \langle \hat{n}_{j} \rangle - 1, \quad  \; j = r, +, -.
\label{man}
\end{eqnarray}
Here, $n_j$ denotes the number of particles in the respective mode which can be uniquely determined from
their associated number operators $\hat{n}_j = \hat{a}^{\dagger}_j \hat{a}_j$. The action of the ladder operators
$\hat{a}^{\dagger}_j$ and $\hat{a}_j$ on the number states $|n_j\rangle$ are given in
equations (\ref{ancr}), (\ref{ops}) and (\ref{op}) respectively.

To evaluate  Mandel's $Q$ parameter we consider the squeezed coherent states be of the form
\begin{eqnarray}
|\xi,\alpha \rangle= N_{\xi, \alpha}\sum_{n=0}^{\infty}\sum_{n_+=0}^{\infty}\sum_{n_{-}=0}^{\infty} c_{n} c_{n_{+},n_{-}} |n, n_{+}, n_{-}\rangle,
\label{scsf2}
\end{eqnarray}
where the constants $c_{n}$ and $c_{n_{+}, n_{-}}$ are given in (\ref{cn}) and (\ref{clm}) with $l\pm m$
should be replaced by $n_{\pm}$.
To find the expectation values $\langle \hat{n}_j \rangle$ and
$\langle \hat{n}^2_j \rangle$, as we did earlier, we first calculate the radial part. Doing so we find

\begin{eqnarray}
\fl \qquad \quad \langle \hat{n}_r \rangle  &=& N^2_{\xi, \alpha} \sum_{n=0}^{\infty} \sum_{n_{+}=0}^{\infty} \sum_{n_{-}=0}^{\infty} c^{*}_{n} c^{*}_{n_{+},n_{-}} c_{n} c_{n_{+},n_{-}} n \label{nrra}\\
\fl \qquad \quad                            &=& \frac{1}{(|\xi_r|^2 - 1)}\left[ |\alpha_r|^2 (1 + |\xi_r|^2 )+ \xi_r \alpha^{*^2}_{r} +
                             \xi^{*}_r \alpha^{2}_{r} + |\xi_r|^2 \right]
\label{nrr}
\end{eqnarray}
and
\small
\begin{eqnarray}
\fl \;\; \langle \hat{n}^2_r \rangle &=& N^2_{\xi, \alpha} \sum_{n=0}^{\infty}\sum_{n_+=0}^{\infty}\sum_{n_{-}=0}^{\infty} c^*_{n} c^*_{n_{+},n_{-}} c_{n} c_{n_{+},n_{-}} n^2 \label{nr2a} \\
\fl \;                          &=& \frac{1}{ (1-|\xi_r|^2)^2 }\left[|\alpha_r|^4 (1 + |\xi_r|^2)^2 +  |\xi_r|^2 (2 + |\xi_r|^2)+(\xi^*_r \alpha^2_r + \xi_r \alpha^{*^2}_{r})^2 \right. \nonumber\\
\fl \;\;  & & \; \left. +  (2 (1 + |\alpha_r|^2 )(1 + |\xi_r|^2) + 2 |\xi_r|^2)(\xi^*_r \alpha^2_r +\xi_r \alpha^{*^2}_{r})  +  |\alpha_r|^2 (1 +  8|\xi_r|^2+ 3|\xi_r|^4 )  \right].
\label{nr2}
\end{eqnarray}
\normalsize
The details of evaluating both the expressions (\ref{nrra}) and (\ref{nr2a}) are given in the \ref{ap2}.

Since $\hat{n}_{\pm}|n_{+}, n_{-}\rangle = n_{\pm}|n_{+}, n_{-}\rangle$, the expectation values
$\langle \hat{n}_{\pm} \rangle$ and $\langle \hat{n}^2_{\pm} \rangle$ associated with the
angular part also provide the same expressions given in (\ref{nrr}) and (\ref{nr2}).
Hence, in general, we can write the expectation values $\langle \hat{n}_j \rangle$ and
$\langle \hat{n}^2_j \rangle$ are  of the form

\small
\begin{eqnarray}
\fl  \quad\langle \hat{n}_j \rangle &=& \frac{1}{(1 -|\xi_j|^2)}\left[ |\alpha_j|^2 (1 + |\xi_j|^2 )+ \xi_j \alpha^{*^2}_{j} +
                             \xi^{*}_j \alpha^{2}_{j} + |\xi_j|^2 \right],
\label{nj}\\
\fl  \quad \langle \hat{n}^2_j \rangle &=& \frac{1}{ (1-|\xi_j|^2)^2 }\left[|\alpha_j|^4 (1 + |\xi_j|^2)^2 +  |\xi_j|^2 (2 + |\xi_j|^2)+(\xi^*_j \alpha^2_j + \xi_j \alpha^{*^2}_{j})^2 \right. \nonumber\\
\fl \;  & & \;\;\; \left. +  (2 (1 + |\alpha_j|^2 )(1 + |\xi_j|^2) + 2 |\xi_j|^2)(\xi^*_j \alpha^2_j +\xi_j \alpha^{*^2}_{j})  +  |\alpha_j|^2 (1 +  8|\xi_j|^2+ 3|\xi_j|^4 )  \right],
\label{nj2}
\end{eqnarray}
\normalsize
where ${\displaystyle \alpha_j = \frac{\alpha_{0,j} - \xi_{j} \alpha^*_{0,j}}{\sqrt{1 - |\xi_j|^2}}}$ and $j = r, +, -$.

\begin{figure}[t]
\centering
\includegraphics[width=0.5\linewidth]{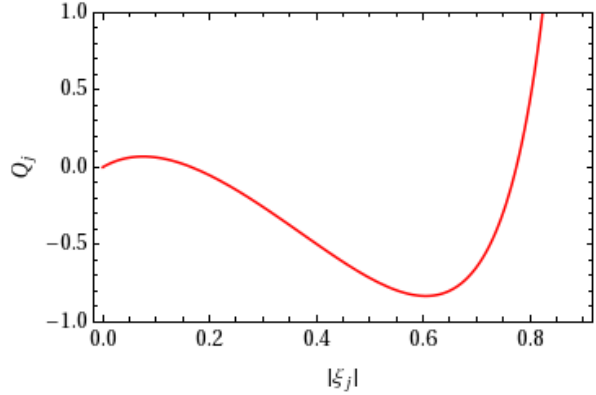}
\vspace{-0.3cm}
\caption{The plot of Mandel's $Q$ parameter for each mode of the states with $\alpha_{0,j} = 3 $.}
\label{fig_man}
\end{figure}

From the expressions (\ref{nj}) and (\ref{nj2}) we can analyze the Poissonian ($Q_{j} = 0)$,
sub-Poissonian $(Q_{j} < 0)$ and super-Poissonian $(Q_{j} > 0)$ nature of the states in each mode.  We
can also calculate the parameter $Q_{j}$  of the three mode squeezed coherent states
which basically depends on the parameters $\alpha_{0, j}$ and $\xi_j$. The obtained numerical results
are depicted in figure \ref{fig_man}. The figure shows that when $|\xi_j| <  1$ and $\alpha_{0,j} = 3$,
the $Q$-parameter takes positive and negative values which in turn confirm the super-Poissonian ($Q_j > 0$) ) and
sub-Poissonian ($Q_j < 0$) nature of the states.
The sub-Poissonian statistics indicates the non-classical nature exhibited by the states.

\subsection{\bf Quadrature squeezing}
The squeezed coherent states given in (\ref{scsf}) are  expressed in terms of the variables
$r, \theta$ and $\phi$. As far as the radial part is concerned one can analyze
the squeezing in the observables such as generalized
position  $(w_r)$ and its conjugate momentum $(p_r)$. To show the squeezing in the angular variables
$\theta$ and $\phi$ we have to consider the angular momentum quantities $L_x, L_y$ and  $L_z$.
In the following, we analyze the squeezing in
position and momentum coordinates which are defined as \cite{walls}
\begin{eqnarray}
\hat{w_r} = \frac{1}{\sqrt{2}} (\hat{a}^{\dagger}_{r} + \hat{a}_r), \qquad \quad \hat{p_r} = \frac{i}{\sqrt{2}} (\hat{a}^{\dagger}_{r} - \hat{a}_r).
\label{qua}
\end{eqnarray}

To analyze the squeezing in the quadratures $\hat{w_r}$ and $\hat{p_r}$ in which
the Heisenberg uncertainty relation holds, $\left(\Delta \hat{w_r} \right)^2\left(\Delta \hat{p_r} \right)^2 \ge \frac{1}{4}$,
where $\Delta \hat{w_r}$ and $\Delta \hat{p_r}$ denote uncertainties in $\hat{w_r}$ and $\hat{p_r}$
respectively, we introduce the following inequalities,
that is
\begin{eqnarray}
\quad I_{1} &=& \langle {\hat{a}_{r}}^2 \rangle + \langle \hat{a}^{\dagger^2}_{r} \rangle - \langle {\hat{a}_{r}} \rangle^{2} - \langle \hat{a}^{\dagger}_{r} \rangle^{2} - 2 \langle \hat{a}_{r} \rangle \langle \hat{a}^{\dagger}_{r} \rangle
+ 2\langle \hat{a}^{\dagger}_{r} \hat{a}_{r} \rangle < 0,
\label{id1}\\
\quad I_{2} &=& -\langle {\hat{a}_{r}}^2 \rangle - \langle \hat{a}^{\dagger^2}_{r} \rangle + \langle {\hat{a}_{r}} \rangle^{2} + \langle \hat{a}^{\dagger}_{r} \rangle^{2} - 2 \langle \hat{a}_{r} \rangle \langle \hat{a}^{\dagger}_{r} \rangle
+ 2\langle \hat{a}^{\dagger}_{r} \hat{a}_{r} \rangle < 0,
\label{id2}
\end{eqnarray}
which can be derived from the squeezing condition
$(\Delta \hat{r})^2 < \frac{1}{2}$ or $(\Delta \hat{p_r})^2 < \frac{1}{2}$ by implementing
the expressions given in (\ref{qua}). The expectation values should be calculated with respect to the squeezed coherent states
$|\xi, \alpha\rangle$ in which the squeezing property has to be examined.

For the squeezed coherent states (\ref{scsf}), we obtain the following
values for the quantities which appear in the equations (\ref{id1}) and (\ref{id2}), that is
\begin{eqnarray}
\fl \quad \quad \langle {\hat{a}_{r}} \rangle &=&  \left(\frac{\alpha_r + |\xi_r| \alpha^{*}_r}{\sqrt{1 - |\xi_r|^2}} \right),
\quad
\langle {\hat{a}^{\dagger}_{r}} \rangle = \left(\frac{\alpha^{*}_r + |\xi_r| \alpha_r}{\sqrt{1 - |\xi_r|^2}} \right),  \label{ar} \\
\fl \quad \quad \langle {\hat{a}^2_{r}} \rangle &=& \frac{\xi_r + (\alpha^{*}_r \xi_r + \alpha_r)^2}{1 - |\xi_r|^2}, \quad
\langle {\hat{a}^{\dagger^2}_{r}} \rangle  = \frac{\xi^{*}_r + (\alpha_r \xi^{*}_r + \alpha^{*}_r)^2}{1 - |\xi_r|^2},
\label{ar2}\\
\fl \quad \quad \langle \hat{a}^{\dagger}_r \hat{a}_{r} \rangle  &=& \frac{1}{(1-|\xi_r|^2)} \left( |\alpha_{r}|^2 (1 + |\xi_r|^2)+ |\xi_r|^2 + \alpha^2_{r}\xi_r+ \alpha^{*^2}_{r}\xi^*_r\right),
\label{arard}
\end{eqnarray}
where ${\displaystyle \alpha_r = \frac{\alpha_{0,r} - \xi_r \alpha^*_{0,r}}{\sqrt{1 - |\xi_r|^2}}}$.
\begin{figure}[h]
\begin{center}
\includegraphics[width=6in]{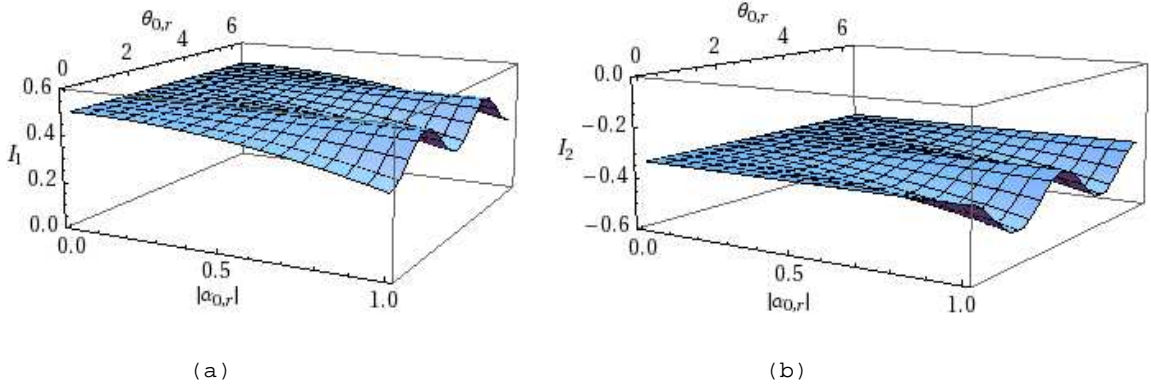}
\end{center}
\caption{The plots of (a) $I_1$ and (b) $I_2$ which are calculated with respect to
squeezed coherent states (\ref{scsf}) with $\xi_r = 0.3$ for different values of $|\alpha_{0,r}|$.}
\label{id_12}
\end{figure}

With the expressions  given in (\ref{ar})-(\ref{arard}) we evaluate the inequalities (\ref{id1}) and (\ref{id2}) numerically and
plot the outcome in figure \ref{id_12} with $\alpha_{0,r} = |\alpha_{0,r}| e^{i \theta_{0,r}} $. From figure \ref{id_12} we
observe that the identities  given in equations (\ref{id1}) and (\ref{id2}) for the squeezed coherent states
$|\xi, \alpha\rangle$, satisfying the uncertainty relation, shows  $I_2 < 0$
and $I_1 > 0$. This in turn confirms the squeezing in
the quadrature $\hat{p}_r$, for all values of $\alpha_{0,r}$.

As far as the angular  part is concerned
the squeezing in the quadratures $\hat{L}_x$ or $\hat{L}_y$ can be analyzed
through either one of the
normalized quantities ${S}_{L_x} < 0$ or ${S}_{L_y} < 0$ where
${S}_{L_x} = \frac{2\Delta \hat{L}^2_{x} - |\langle \hat{L}_{z} \rangle|}{|\langle \hat{L}_{z} \rangle|}$ and
${{S}_{L_y}} = \frac{2 \Delta \hat{L}^2_{y} -|\langle \hat{L}_{z} \rangle| }{|\langle \hat{L}_{z} \rangle|}$ \cite{ang}.
We can express the operators $\hat{L}_{x}$ and $\hat{L}_{y}$ in
terms of $\hat{L}_{+}$ and $\hat{L}_{-}$, namely
\begin{eqnarray}
\hat{L}_{x} = \frac{1}{2} \left(\hat{L}_{+} + \hat{L}_{-}\right), \qquad \hat{L}_{y} = \frac{i}{2} \left(\hat{L}_{+} - \hat{L}_{-}\right)
\end{eqnarray}
so that the uncertainty of $\hat{L}_{x}$ and $\hat{L}_y$ can now be expressed in terms of $\hat{L}_{+}$ and $\hat{L}_{-}$, that is
\begin{eqnarray}
\fl \qquad (\Delta \hat{L}_{x})^2 &=& \langle \hat{L}^2_{x} \rangle - \langle \hat{L}_{x} \rangle^2 \nonumber \\
 \fl \qquad                      &=& \frac{1}{4} \left(\langle \hat{L}^2_{+} \rangle + \langle \hat{L}^2_{-} \rangle + \langle \hat{L}_{+}\hat{L}_{-}\rangle +  \langle \hat{L}_{-}\hat{L}_{+}\rangle + (\langle \hat{L}_{+}\rangle + \langle \hat{L}_{-}\rangle )^2\right),
\label{ulx}\\
\fl \qquad(\Delta \hat{L}_{y})^2 &=& \langle \hat{L}^2_{y} \rangle - \langle \hat{L}_{y} \rangle^2 \nonumber \\
 \fl \qquad                      &=& -\frac{1}{4} \left(\langle \hat{L}^2_{+} \rangle + \langle \hat{L}^2_{-} \rangle - \langle \hat{L}_{+}\hat{L}_{-}\rangle -  \langle \hat{L}_{-}\hat{L}_{+}\rangle - (\langle \hat{L}_{+}\rangle - \langle \hat{L}_{-}\rangle )^2\right).
\label{uly}
 \end{eqnarray}

\begin{figure}[t]
\centering
\includegraphics[width=0.9\linewidth]{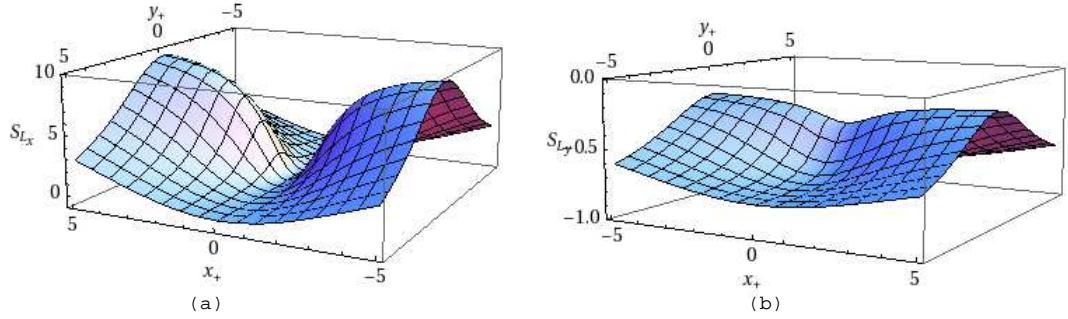}
\vspace{-0.1cm}
\caption{The plot of normalized quantities (a) ${S}_{L_x}$ and (b)${S}_{L_y}$ with
$\xi_{+} = \xi_{-} = 0.1$ and $\alpha_{0,-} = 1.3$.}
\label{sqa}
\end{figure}

For the squeezed coherent states (\ref{scsf}), the expectation values in (\ref{ulx}) and (\ref{uly}) are found to be
\small
\begin{eqnarray}
\fl \qquad \left\langle \hat{L}_+\right\rangle &=& \left(\frac{\alpha^*_+- \xi^{*}_+ \alpha_+}{\sqrt{1 - |\xi_+|^2}}\right)
                                      \left(\frac{\alpha_-- \xi_- \alpha^*_+}{\sqrt{1 - |\xi_-|^2}} \label{lp}\right),\label{lp}\\
\fl \qquad\left\langle \hat{L}_-\right\rangle &=& \left( \frac{\alpha_+- \xi_+ \alpha^{*}_+}{\sqrt{1 - |\xi_+|^2}} \right)
                                      \left(\frac{\alpha^*_{-}- \xi^*_{-} \alpha_{+}}{\sqrt{1 - |\xi_-|^2}}\right), \label{lm}\\
\fl \qquad\left\langle \hat{L}_+^2\right\rangle &=& \frac{\xi _+^*\xi _-}{\sqrt{1 - \left|\xi _+\right|^2}\sqrt{1 -\left|\xi _-\right|^2}} \left(\frac{\left(\alpha _+^* - \xi _+^*\alpha _+\right)^2}{\xi _+} + 1\right)\left(\frac{\left(\alpha _{-}-
\xi _- \alpha _-^*\right)^2}{\xi _-^*} + 1\right), \label{lp2} \\
\fl \qquad \left\langle \hat{L}_-^2\right\rangle &=& \frac{\xi _-^*\xi _+}{\sqrt{1 - \left|\xi _+\right|^2}\sqrt{1 -\left|\xi _-\right|^2}} \left( \frac{ \left(\alpha _+-\xi _+\alpha _+^*\right)^2}{\xi _+^*} + 1\right)\left( \frac{\left(\alpha _-^*-\xi _-^*
\alpha _-\right)^2}{\xi _-} + 1\right), \label{lm2}\\
\fl \qquad \left\langle \hat{L}_{+}\hat{L}_{-}\right\rangle &=& \left( \left|\frac{\alpha _+^*-\xi _+^* \alpha _+}{\sqrt{\left|\xi _+\right|^2 - 1}}\right|^2 + \frac{\left|\xi _+\right|^2}{\left|\xi _+\right|^2 - 1}\right)\left( 1 +\left|\frac{\alpha _-^*-\xi _-^*\alpha _-}{\sqrt{\left|\xi _-\right|^2 - 1}}\right|^2 + \frac{\left|\xi _-\right|^2}{\left|\xi _-\right|^2 - 1} \right), \label{lpm}\\
\fl \qquad \left\langle \hat{L}_{-}\hat{L}_{+}\right\rangle &=& \left( \left|\frac{\alpha _+-\xi _+ \alpha _+}{\sqrt{\left|\xi _+\right|^2 - 1}}\right|^2 + \frac{\left|\xi _+\right|^2}{\left|\xi _+\right|^2 - 1} + 1\right)\left( \left|\frac{\alpha _--\xi _-\alpha _-}{\sqrt{\left|\xi _-\right|^2 - 1}}\right|^2 + \frac{\left|\xi _-\right|^2}{\left|\xi _-\right|^2 - 1} \right), \label{lmp}\\
\fl \qquad\left\langle \hat{L}_z\right\rangle &=& \frac{1}{2 \left(\left|\xi _+\right|^2 - 1\right)}\left(\left(\alpha _+-\xi _+\alpha_+^*\right)\left(\alpha_+^*-\xi_+^* \alpha_+\right) +\left|\xi_+\right| \right)- \frac{1}{2 \left(\left|\xi_-\right|^2 - 1\right)} \nonumber \\
\fl \qquad & & \times \left(\left(\alpha_{-} - \xi_{-}\alpha_{-}^{*}\right)\left(\alpha_{-}^{*}-\xi_{-}^{*} \alpha_{-}\right) +\left|\xi_{-}\right|\right)
\label{lz},
\end{eqnarray}
\normalsize
where ${\displaystyle \alpha_{\pm} = \frac{\alpha_{0,\pm} - \xi_{\pm} \alpha^*_{0,\pm}}{\sqrt{1 - |\xi_{\pm}|^2}}}$.

We evaluate the normalized quantities ${S}_{L_x}$ and ${S}_{L_y}$ numerically
by substituting the expectation values (\ref{lp})-(\ref{lz}) in the uncertainties
(\ref{ulx}) and (\ref{uly}). We plot the numerical
results in figure \ref{sqa} where we have considered $\alpha_{0,+} = x_{+} + i y_{+}$ with
$\xi_{+} = \xi_{-} = 0.1$ and $\alpha_{0,-} = 1.3$. For this choice of parameters
figure \ref{sqa} shows that $S_{L_x} > 0$
and $S_{L_y} <0$  which explicitly demonstrates the squeezing in $\hat{L}_{y}$.

\section{\bf Wigner function for $3$-dimensional isotonic oscillator}
\label{sec3}
In this sub-section, we evaluate Wigner function $(W(x, p))$ of the squeezed coherent states (\ref{scs}).
The Wigner function (a quasi-probability distribution function),
which was introduced as quantum corrections in classical statistical mechanics, normally takes negative values in
certain domains of phase space so that it cannot be interpreted as a classical distribution function which is non-negative
by necessity \cite{vbook}.

The Wigner function for the single mode state is  described by the density
operator $\hat{\rho}$ which can be written as \cite{spara}
\begin{eqnarray}
\fl \quad \qquad W(\zeta) = 2\;\mbox{Tr}[\hat{\rho} \hat{T}(\zeta, 0)], \qquad
\hat{T}(\zeta, 0) = \hat{D}(\zeta, \zeta^{*}) e^{{i}\pi \hat{a}^{\dagger} \hat{a}}  \hat{D}^{-1}(\zeta, \zeta^{*}),
\label{wig}
\end{eqnarray}
where $\hat{D}(\zeta, \zeta^{*})$ and $\hat{\rho}$ are the displacement and
density operators respectively and $\hat{T}$ is the complex Fourier transform of the $s$-parameterized
displacement operator, $\hat{D}(\eta, s)$, in which  $\eta$ is the coherent eigenvalue with
$s = 0$ corresponding to the Wigner distribution function.

Since  $\hat{\rho} = |\xi, \alpha\rangle \langle \xi, \alpha|$, the Wigner function for a single mode
can be explicitly calculated from
\begin{eqnarray}
\qquad \quad W(\zeta) =  2 \langle \xi, \alpha|\hat{T}(\zeta, 0)|\alpha, \xi \rangle.  \label{w3}
\end{eqnarray}

Prolonging the definition (\ref{w3}) to three orthogonal modes \cite{spara_N}, we get
\begin{eqnarray}
\fl \quad \qquad W(\{\zeta_{1}, \zeta_{2}, \zeta_{3} \}) = 8\;\mbox{Tr}[\hat{\rho} \prod^{3}_{i = 1} {\hat{T}}_{i}(\zeta_{i}, 0)], \qquad
{\hat{T}}_{i}(\zeta_{i}, 0) = \hat{D}(\zeta_i) e^{i\;\pi \hat{a}^{\dagger}_{i} \hat{a}_i} \hat{D}^{-1}(\zeta_i),
\label{w32}
\end{eqnarray}
where $\zeta_{1} = \zeta_{r},\; \zeta_{2} = \zeta_{+}$ and $\zeta_{3} = \zeta_{-}$.
By applying this definition, (\ref{w32}), to the three mode squeezed coherent states (\ref{scs}), we obtain
\begin{eqnarray}
\fl \quad\;  W(\{\zeta_{r}, \zeta_{+}, \zeta_{-}\})= N^{2}_{\xi,\alpha} \sum^{\infty}_{n, n' = 0} \sum^{\infty}_{l, l' = 0} \sum^{l}_{m = -l} \sum^{l'}_{m' = -l'}
c^{*}_{n'} c_{n} c^{*}_{l',m'} c_{l,m} \nonumber \\ \quad \qquad\times  \langle l', m'| \langle n' | {\hat{T}}(\zeta_{r}, 0) \hat{T}(\zeta_{+}, 0) {\hat{T}}(\zeta_{-}, 0) |n\rangle |l, m\rangle,
\label{w4}
\end{eqnarray}
where $c_{n}$ and $c_{l, m}$ are given in (\ref{cn}) and (\ref{clm}). The  above equation can be evaluated
by separating  the radial part from angular part, that is
\begin{eqnarray}
\fl \qquad\;  W(\{\zeta_{r}, \zeta_{+}, \zeta_{-}\})   =  {N}^2_{\xi,\alpha}  \sum^{\infty}_{n, n' = 0} \sum^{\infty}_{l, l' = 0} \sum^{l}_{m = -l} \sum^{l'}_{m = -l'}
c^{*}_{n'} c_{n} c^{*}_{l',m'} c_{l,m} \langle n' | {\hat{T}}(\zeta_{r}, 0) |n\rangle \nonumber \\ \qquad \qquad \qquad
\times \langle l', m'|{\hat{T}}(\zeta_{+}, 0) \hat{T}(\zeta_{-}, 0) |l, m\rangle. \label{w5}
\end{eqnarray}

The operation of the dual state $\langle n'|$ on the state $\hat{T}(\zeta_{r}, 0)|n\rangle$ may be known as transition 
probability,
expressed in terms of $n$, has already been reported in Ref. \cite{spara}.  The result shows that
\begin{eqnarray}
\fl \qquad \langle n'|{\hat{T}}(\zeta_{r}, 0) |n\rangle  = e^{-2|\zeta_{r}|^2}\left(\frac{n'!}{n!}\right)^{1/2} 2^{n - n' + 1}
(-1)^{n'} (\zeta_{r}^{*})^{n-n'} L^{n-n'}_{n'}(4 |\zeta_{r}|^2),
\label{w6}
\end{eqnarray}
where $L^{n - n'}_{n}$ is the associated Laguerre polynomial of degree $n$.  In a similar way,
the quantities corresponding to the operators $\hat{T}(\zeta_{+}, 0)$
 and $\hat{T}(\zeta_{-}, 0)$  can also be found  in terms of
the number states $n_{\pm}$.  The resultant expressions are turned out to be
\begin{eqnarray}
\fl \quad \langle n'_{+} |  {\hat{T}}(\zeta_{+}, 0) |n_{+}\rangle  = e^{-2|\zeta_{+}|^2}\left(\frac{n'_{+}!}{n_{+}!}\right)^{1/2} 2^{n_{+} - n'_{+} + 1}
(-1)^{n'_{+}} (\zeta_{+}^{*})^{n_{+}-n'_{+}} L^{n_{+}-n'_{+}}_{n'_{+}}(4 |\zeta_{r}|^2),
\label{w6a}
\end{eqnarray}
and
\begin{eqnarray}
\fl \quad \langle n'_{-} |{\hat{T}}(\zeta_{-}, 0) |n_{-}\rangle  = e^{-2|\zeta_{-}|^2}\left(\frac{n'_{-}!}{n_{-}!}\right)^{1/2} 2^{n_{-} - n'_{-} + 1}
(-1)^{n'_{-}} (\zeta_{-}^{*})^{n_{-}-n'_{-}} L^{n_{-}-n'_{-}}_{n'_{-}}(4 |\zeta_{-}|^2).
\label{w6b}
\end{eqnarray}
Substituting (\ref{w6})-(\ref{w6b}) in (\ref{w5}), we can
obtain  Wigner function for the squeezed coherent states (\ref{scs}) with $n_+ = l + m, n'_{+} = l' + m' $ and
$n_- = l - m, n'_{-} = l' - m' $ in the form
\begin{eqnarray}
\fl \; W(\{\zeta_{r}, \zeta_{+}, \zeta_{-}\}) = e^{-2|\zeta_{r}|^2-2|\zeta_{+}|^2-2|\zeta_{-}|^2} {N}^2_{\xi,\alpha}\sum^{\infty}_{n, n' = 0} \sum^{\infty}_{l, l' = 0} \sum^{l}_{m = -l} \sum^{l'}_{m' = -l'} c^{*}_{n'} c_{n} c^{*}_{l',m'} c_{l,m} 2^{n + 2 l - n' - 2 l'+ 3} \nonumber \\
\;\;\;\;\;\times \;\left(\frac{n'! (l'+m')!(l'-m')!}{n!(l+m)!(l-m)!}\right)^{1/2}
(-1)^{n'} (\zeta^{*}_{r})^{n -n'} (\zeta^{*}_{+})^{l + m - l' - m'}(\zeta^{*}_{-})^{l - m - l' + m' }\nonumber \\
\;\;\;\;\; \times L^{n-n'}_{n'}(4 |\zeta_{r}|^2)L^{l+m-l'-m'}_{l'+m'}(4 |\zeta_{+}|^2)L^{l-m-l'+m'}_{l'-m'}(4 |\zeta_{-}|^2),
\label{wigf}
\end{eqnarray}
where $\zeta_{j} = x_{j} + p_{j}$, $ \xi_{i} = -\frac{1}{2} \tanh{R}_j e^{i \phi}_{j}$
and $\alpha_{j} = \alpha_{0,j}\cosh{R_{i}} +
\alpha^{*}_{0,i} e^{i \phi_{j}} \sinh{R_{j}} $, $j = r, \pm $.

On evaluating the right hand side (the details are given in the Appendix C), we find
\begin{eqnarray}
\fl \quad W(\{\zeta_r, \zeta_+, \zeta_-\}) &=& \Pi^{3}_{j=1}\exp{\left[\frac{-2|\zeta_{j}|^2 (|\xi_{j}|^2 + 1) + 2 (\xi_{j} \zeta^{*^2}_{j} + \xi^*_j \zeta^2_{j})}{1 - |\xi_{j}|^2} - 2 |\alpha_{j}|^2   \right. }\nonumber \\
\fl \quad       & & \qquad \qquad + \frac{2(\alpha_j \zeta^{*}_j + \alpha^{*}_j \zeta_j)}{\sqrt{1 - |\xi_j|^2}}
 {\left. - \frac{2  (\xi^{*}_j\alpha_r\zeta_j+ \xi_r\alpha^*_{j} \zeta^{*}_j)}{\sqrt{1-|\xi_j|^2}}\right]}.
\label{re_wig}
\end{eqnarray}

We analyze the Wigner function graphically by considering
$\zeta_{r}=\zeta_{+} = \zeta_{-}$,   $\xi_{r}=\xi_{+} = \xi_{-}$ and $\alpha_{0,r}=\alpha_{0,+} = \alpha_{0,-}$.
With this assumption, the Wigner function of the three mode states can be seen as one mode state.
We then calculate the function $W(x, p)$ numerically and plot the outcome in figure \ref{wig} for
a specific value of  $R, \phi $ and $\alpha_{0}$ with $\zeta =  x + i\;p$.

\begin{figure}[htP]
\centering
\includegraphics[width=0.45\linewidth]{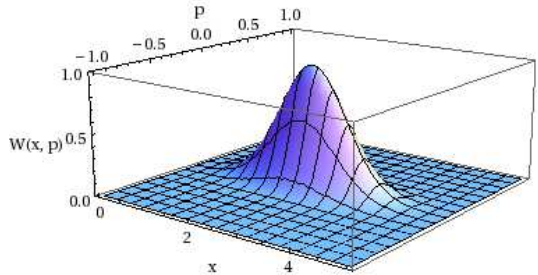}
\vspace{-0.1cm}
\caption{The plot of Wigner function corresponding to
squeezed coherent states (\ref{scs}) with  (a) $R = 0.7 , \phi = 0$ and $\alpha_{0} = 0.5$} \label{wig}
\end{figure}

The plot confirms that the Wigner function
takes smooth shape in which the squeezing effect can be seen.

\section{Conclusion}
In this paper, we have constructed squeezed coherent states and studied their
nonclassical properties associated with the $3$-dimensional generalized isotonic oscillator.
Since the potential under consideration is  a spherically
symmetric one, we  dealt the radial and angular momentum parts separately.
As far as the radial part is concerned we observed that it is
nothing but the newly found extended radial oscillator whose eigenfunctions are
expressed in terms of the recently discovered $X_1$-Laguerre polynomials. We have shown that
this radial part exhibits shape invariance property. Using this property
and implementing Balantekin's method, we have obtained two new operators
which in turn constitute the intertwining supersymmetric operators.  We have demonstrated
that these operators produce  Heisenberg-Weyl algebra and perfectly annihilate and create the eigenstates
of the radial part. Using these ladder operators  we  have
constructed squeezed coherent states of the radial part.
As far as the angular part is concerned we have used Schwinger's representation
to define the creation and annihilation operators. We have derived the associated squeezed coherent states
by considering the disentangled form of two mode squeezing
and displacement operators with certain restrictions on the
allowed values of angular momentum variables $l$ and $m$.
Finally, we have expressed the three mode squeezed coherent states of the
generalized isotonic oscillator as  a tensor product of the squeezed coherent states of the
radial part and squeezed coherent states of the  angular momentum part. To make our results more rigorous we have
proved that these three mode squeezed coherent states resolve the identity operator.
We have explicitly illustrated  the squeezing properties of the constructed
$3$-dimensional squeezed states both in the radial part and in the angular part.
As far as the radial part is concerned we have considered
the conjugate variables, generalized position $w_r$ and the momentum $p_r$
and proved the quadrature squeezing. Similarly, to show the squeezing
in the angular momentum we have considered two normalized quantities,
 namely $S_{L_x}= \frac{2(\Delta \hat{L}_{x})^2 - |\langle \hat{L}_{z} \rangle|}{|\langle \hat{L}_{z} \rangle|}$
and $S_{L_y}=  \frac{2 (\Delta \hat{L}_{y})^2 - |\langle \hat{L}_{z} \rangle|}{|\langle \hat{L}_{z} \rangle|}$ and
demonstrated explicitly the squeezing properties possessed by these states in
the angular variables.
In addition to the above, we have
evaluated Wigner function of these squeezed coherent states and shown that these
states exhibit squeezing property. Since the ladder operators of this $3$-dimensional system
act on the eigenstates only linearly, one can deform these operators nonlinearly and construct certain
non-classical states such as nonlinear coherent states, nonlinear squeezed states, generalized intelligent states
and so on. The details will be presented elsewhere.

\section*{Acknowledgment}
VC  wishes to thank the Council of Scientific and Industrial Research,
Government of India, for providing a Senior Research Fellowship.

\begin{appendix}

\section{\bf Evaluation of (\ref{ope}) and (\ref{ope2})}
\label{appa}

To evaluate the terms inside the square bracket in (\ref{ope}) we recall the following
property associated with the $X_{1}$- Laguerre polynomials \cite{kam}. The differential equation
(\ref{diff}) can be factorized by two operators \cite{kam}, namely
\begin{eqnarray}
A_k(y) &=& - \frac{(x + k + 1)^2}{(x+k)} \frac{d}{dx}\left(\frac{y}{x+k+1}\right), \label{ak}\\
\fl \mbox{and}\nonumber \\
B_k(y) &=& \frac{x(x + k)}{(x+k+1)} (y{'} - y) + k y.
\label{bk}
\end{eqnarray}
One can unambiguously check $A_{k} B_{k} (y)$ factorizes the equation (\ref{diff}) as
$A_{k} B_{k}(y) = \lambda (y)$.

Rescaling the variable $x=\omega r^2$ and the constant  $k=l+\frac{1}{2}$ in (\ref{bk}), we get
\begin{eqnarray}
\fl \; B_l(y)= -\frac{r}{2} \left(\frac{2\omega r^2 + 2 l + 1}{2\omega r^2 + 2 l + 3}\right) \left[ -\frac{dy}{dr} +\left(
2 \omega r - \frac{2 l + 3}{r} - \frac{4 \omega r}{(2 \omega r^2 + 2 l + 1)}\right)y\right].
\label{bka}
\end{eqnarray}

Using the identity \cite{kam} $B_l \hat{L}^{\left(l + \frac{3}{2}\right)}_{n} = n \hat{L}^{\left(l + \frac{1}{2}\right)}_{n+1}$
with $y = \hat{L}^{\left(l+ \frac{3}{2}\right)}_{n+1}$, equation (\ref{bka}) can be brought to the form
\begin{eqnarray}
\fl \qquad \quad -\frac{d \hat{L}^{\left(l+ \frac{3}{2}\right)}_{n+1}}{dr} + \left(2 \omega r - \frac{2 l + 3}{r} + \frac{4 \omega r}{(2\omega r^2 + 2l + 3)}  \right) \hat{L}^{\left(l+\frac{3}{2}\right)}_{n+1} =  -\frac{2}{r}\left(\frac{2\omega r^2 + 2 l + 3}{2\omega r^2 + 2 l + 1}\right)\; \nonumber \\ \hspace{9cm} \times (n+1)\; \hat{L}^{\left(l+\frac{1}{2}\right)}_{n+2}.
\label{bkx}
\end{eqnarray}
We use this expression to simplify the equation (\ref{ope}).

To evaluate (\ref{ope2}), we again rescale the relation (\ref{ak}) in such
a way that $x= \omega r^2$ and $k=l+\frac{1}{2}$ so that the identity (\ref{ak}) now becomes
\begin{eqnarray}
\fl \qquad \quad A_l(y)= -\frac{1}{2 \omega r} \left(\frac{2\omega r^2 + 2 l + 3}{2\omega r^2 + 2 l + 1}\right)
\left[ \frac{dy}{dr} - \frac{4 \omega r}{(2 \omega r^2 + 2 l + 3) }y\right].
\label{ak2}
\end{eqnarray}
Recalling another identity $A_l \hat{L}^{\left(l + \frac{1}{2}\right)}_{n+1}= \hat{L}^{\left(l + \frac{3}{2}\right)}_{n}$
with $y = \hat{L}^{\left(l + \frac{1}{2}\right)}_{n+1}$, equation (\ref{ak2}) can be rewritten as
\begin{eqnarray}
\fl \qquad \quad \frac{d\hat{L}^{\left(l + \frac{1}{2}\right)}_{n+1}}{dr} - \frac{4 \omega r}{(2 \omega r^2 + 2 l + 3) } \hat{L}^{\left(l + \frac{1}{2}\right)}_{n+1} = -2 \omega r\left(\frac{2\omega r^2 + 2 l + 1}{2\omega r^2 + 2 l + 3}\right)\; \hat{L}^{\left(l+\frac{3}{2}\right)}_{n}.
\end{eqnarray}

This relation is used to simplify the intertwining relation  (\ref{ope2}).

\section{\bf Evaluation of (\ref{nrra}) and (\ref{nr2a})}
\label{ap2}
In the following, we discuss the method of evaluating the series $\langle \hat{n}_r \rangle$ given in (\ref{nrra}),
that is
\begin{eqnarray}
\langle \hat{n}_r \rangle = N^{2}_{\xi, \alpha} \sum^{\infty}_{n = 0}\sum^{\infty}_{n_{+} = 0} \sum^{\infty}_{n_{-} = 0}
 c^{*}_{n} c^{*}_{n_{+},n_{-}} c_{n} c_{n_{+},n_{-}} n. 
\nonumber  \hspace{4.5cm} \mbox{(\ref{nrra})}
\end{eqnarray}
With the definition  $N^{-2}_{\pm} = \sum^{\infty}_{n_{+} = 0} \sum^{\infty}_{n_{-} = 0}
c^{*}_{n_{+},n_{-}} c_{n_{+},n_{-}}$, where $N^{2}_{\pm}$ is given in  (\ref{npm}),
the above equation (\ref{nrra}) can be written in  the following compact form
\begin{eqnarray}
\qquad \langle \hat{n}_r \rangle = N^{2}_{r} \sum^{\infty}_{n = 0} c^{*}_{n} c_{n} n.
\label{a2.2}
\end{eqnarray}
Substituting (\ref{cn}) in  (\ref{a2.2}) and redefining the summation appropriately, we obtain
\begin{eqnarray}
\fl \qquad \qquad \langle \hat{n}_r \rangle = N^{2}_{r} \sum^{\infty}_{s = 0} \frac{H_{s+1}(x^{*}_r)H_{s+1}(x_r)}{s!}\left(\frac{-|\xi_r|}{2}\right)^{s+1}.
\label{a2.3}
\end{eqnarray}

Recalling the integral representation associated with the
Hermite polynomials, namely $H_{n}(z) = \frac{2^n}{\sqrt{\pi}}\int^{-\infty}_{\infty} e^{-t^2} (z + i t)^n dt$ \cite{book},
the above equation  (\ref{a2.3}) can be brought to the form
 \begin{eqnarray}
\fl \qquad \qquad \langle \hat{n}_r \rangle &=& \frac{-2|\xi_r|N^{2}_{r}}{\pi}\int^{\infty}_{-\infty}\int^{\infty}_{-\infty} e^{-z^2_1-z^2_2}
(x_r + i z_1)(x^*_r + i z_2)\nonumber \\
 \fl \qquad \qquad & &\qquad \qquad \times \sum^{\infty}_{s = 0} \frac{(- 2 |\xi_r|(x_r + i z_1)(x^*_r + i z_2))^s}{s!} dz_1 dz_2.
\label{a2.4}
\end{eqnarray}

We observe that the summation inside the integral can also be written as an exponential function.
The non-exponential terms left over inside
the integrals, that is $(x_r + iz_1) (x^*_r + i z _2)$ can also be brought into  exponential form through the identity
$ (x_r + i z_1)(x^*_r + i z_2) = \left\{\frac{d}{dp}\left[ \exp{\left[ p (x_r + i z_1)(x^*_r + i z_2)\right]}\right]\right\}_{p = 0}$. As a result equation (\ref{a2.4})  now becomes
\small
 \begin{eqnarray}
\fl \; \langle \hat{n}_r \rangle = \frac{-2|\xi_r|N^{2}_{r}}{\pi} \nonumber \\
\fl \qquad \qquad  \times\left\{\frac{d}{dp}\left[\int^{\infty}_{-\infty}\int^{\infty}_{-\infty} \exp{\left[-z^2_1-z^2_2 + (p- 2 |\xi_r|)(x_r + i z_1)(x^*_r + i z_2)\right]} dz_1 dz_2\right]\right\}_{p =0}.
\label{a2.5}
\end{eqnarray}
\normalsize
To evaluate (\ref{a2.5}) we introduce the transformation $\eta = p  - 2 |\xi_r|$. In the new variable
equation (\ref{a2.5}) reads
\small
\begin{eqnarray}
\fl \; \langle \hat{n}_r \rangle = \frac{-2|\xi_r|N^{2}_{r}}{\pi}\left\{\frac{d}{d\eta}\left[\int^{\infty}_{-\infty}\int^{\infty}_{-\infty} \exp{\left[-z^2_1-z^2_2 + \eta (x_r + i z_1)(x^*_r + i z_2)\right]} dz_1 dz_2\right]\right\}_{\eta = - 2 |\xi_r|}.
\label{a2.5bb}
\end{eqnarray}
\normalsize

To begin with we evaluate the following integral in (\ref{a2.5bb}), namely
\begin{eqnarray}
I_1 = \int^{\infty}_{-\infty}\exp{\left[-z^2_1 + \eta (x_r + i z_1)(x^*_r + i z_2)\right]} dz_1.
\end{eqnarray}
We noted that this integral is nothing but the  Fourier transform of an exponential function. With
this identification we find
\begin{eqnarray}
I_1 =\sqrt{\frac{\pi}{1 - \frac{\eta^2}{4}}}\exp{\left[\frac{\eta^2}{4}z^2_2-\frac{x^{*^2}_r \eta^2}{4} - i \frac{\eta^2}{2} x^*_r z_2+ \eta |x_r|^2 + i \eta x_r z_2\right]}.
\label{a2.6a}
\end{eqnarray}
Substituting (\ref{a2.6a}) in the second integral in (\ref{a2.5bb}), we
get
\begin{eqnarray}
I_2 = \int^{\infty}_{-\infty}\exp{\left[-z^2_2\right]} I_1 dz_2.
\label{a2.6b}
\end{eqnarray}
The above integral also turns out to be the  Fourier transform of an exponential function. 
The explicit integration leads us to
\begin{eqnarray}
I_2 = \frac{\pi}{\sqrt{1 - \frac{\eta^2}{4}}}\exp{\left[\eta|x_r|^2-\frac{x^{*^2}_r \eta^2}{4}-\frac{\eta^2}{4\left(1 - \frac{\eta^2}{4}\right)}\left(x_r - \frac{\eta}{2} x^*_r \right)^2 \right]}.
\label{a2.6d}
\end{eqnarray}

Substituting (\ref{a2.6d}) in (\ref{a2.5bb}) and simplifying the resultant expression we arrive at
\begin{eqnarray}
\fl \quad \qquad \langle \hat{n}_r \rangle = -2|\xi_r|N^{2}_{r}\left\{\frac{d}{d\eta}\left[
\frac{1}{\sqrt{1 - \frac{\eta^2}{4}}}\exp{\left[\frac{\eta|x_r|^2 - \frac{\eta^2}{4}(x^2_r + x^{*^2}_r)}{1 - \frac{\eta^2}{4}}\right]}\right]\right\}_{\eta = - 2 |\xi_r|}.
\label{a2.6}
\end{eqnarray}

Now carrying out the differentiation and substituting the expressions
$x_r \left(= \frac{\alpha_r \sqrt{1 - |\xi_r|^2}}{\sqrt{-2 \xi_r}}\right)$
and its complex conjugate in the resultant equation we obtain the
following  expression for $\langle \hat{n}_r \rangle$, namely
\begin{eqnarray}
\fl \qquad \qquad \langle \hat{n}_r \rangle = \frac{1}{(1 - |\xi_r|^2)} \left[|\alpha_r|^2 (1 + |\xi_r|^2) + \alpha^2 \xi^{*}_r + \alpha^{*^2}_{r} \xi_r + |\xi_r|^2 \right]. \nonumber
 \hspace{2.5cm} \mbox{(\ref{nrr})}
\end{eqnarray}
We use this expression to evaluate the Mandel's $Q$ parameter.

Now we evaluate the expectation value  $\langle \hat{n}^2_r \rangle$  given in (\ref{nr2a}), that is
\begin{eqnarray}
\langle \hat{n}^2_r \rangle = N^{2}_{\xi, \alpha} \sum^{\infty}_{n = 0}\sum^{\infty}_{n_{+} = 0} \sum^{\infty}_{n_{-} = 0}
 c^{*}_{n} c^{*}_{n_{+},n_{-}} c_{n} c_{n_{+},n_{-}} n^2.
\label{a3.1}
\end{eqnarray}
Separating the radial part from angular part by
recalling $N^{-2}_{\pm} = \sum^{\infty}_{n_{+} = 0} \sum^{\infty}_{n_{-} = 0}
c^{*}_{n_{+},n_{-}} c_{n_{+},n_{-}}$, with $N^{2}_{\pm}$ defined in (\ref{npm}),
the triple sum in (\ref{a3.1}) can be reduced to the form
\begin{eqnarray}
\qquad \langle \hat{n}^2_r \rangle = N^{2}_{r} \sum^{\infty}_{n = 0} c^{*}_{n} c_{n} n^2.
\label{a3.2}
\end{eqnarray}
Substituting (\ref{cn}) in  (\ref{a3.2}) and redefining the summation appropriately, we
find
\begin{eqnarray}
\fl \quad \quad \quad \quad \langle \hat{n}^2_r \rangle &=& N^{2}_{r} \sum^{\infty}_{s = 0} \frac{H_{s+2}(x^{*}_r)H_{s+2}(x_r)}{s!}\left(\frac{-|\xi_r|}{2}\right)^{s+2} \nonumber \\
\fl \quad \quad \quad \quad & &  \qquad \qquad \quad+  N^{2}_{r} \sum^{\infty}_{j = 0} \frac{H_{j+1}(x^{*}_r)H_{j+1}(x_r)}{j!}\left(\frac{-|\xi_r|}{2}\right)^{j+1}.
\label{a3.3}
\end{eqnarray}

The second term in (\ref{a3.3}) is
nothing but $\langle \hat{n}_r \rangle$ (vide equation (\ref{a2.3}))  which has already been evaluated
(vide equation (\ref{nrr})). As a consequence, we confine our attention only on the first term in
equation (\ref{a3.3}) (which we call as $G$), namely
\begin{eqnarray}
\fl \qquad \qquad \qquad G = N^{2}_{r} \sum^{\infty}_{s = 0} \frac{H_{s+2}(x^{*}_r)H_{s+2}(x_r)}{s!}\left(\frac{-|\xi_r|}{2}\right)^{s+2}.
\label{a3.4}
\end{eqnarray}

Using the integral representation,
$H_{n}(z) = \frac{2^n}{\sqrt{\pi}}\int^{\infty}_{-\infty} e^{-t^2} (z + i t)^n dt$ , we can rewrite  (\ref{a3.4}) as

 \begin{eqnarray}
\fl \qquad \qquad  G &=& \frac{4N^{2}_{r} |\xi_r|^2}{\pi}\int^{\infty}_{-\infty}\int^{\infty}_{-\infty} e^{-z^2_1-z^2_2}
(x_r + i z_1)^2(x^*_r + i z_2)^2 \nonumber \\
                     & &\qquad \qquad \times \sum^{\infty}_{s = 0} \frac{(- 2 |\xi_r|(x_r + i z_1)(x^*_r + i z_2))^s}{s!} dz_1 dz_2.
\label{a3.5}
\end{eqnarray}

By recognizing the sum appearing in (\ref{a3.5}) is nothing but an exponential function and the terms
$(x_r + i z_1)^2 (x^{*}_r + i z_2)^2$  can also be rewritten as an exponential function
we can bring equation to the form
\small
\begin{eqnarray}
\fl \;\;\;   G = \frac{4N^{2}_{r} |\xi_r|^2}{\pi} \nonumber \\
\fl \qquad \quad  \times \left\{\frac{d^2}{dp^2}\left[\int^{\infty}_{-\infty}\int^{\infty}_{-\infty} \exp{\left[-z^2_1-z^2_2  + (p- 2 |\xi_r|)(x_r + i z_1)(x_r + i z_2)\right]} dz_1 dz_2\right]\right\}_{p =0}.
\label{a3.6}
\end{eqnarray}
\normalsize

To evaluate the integrals we introduce the transformation $\eta = p  - 2 |\xi_r|$  so that the
equation (\ref{a3.6}) in the new variable reads
\begin{eqnarray}
\fl \;\;\; G = \frac{4N^{2}_{r} |\xi_r|^2}{\pi}\nonumber \\
\fl \qquad \;\;\times \left\{\frac{d^2}{d\eta^2}\left[\int^{\infty}_{-\infty}\int^{\infty}_{-\infty} \exp{\left[-z^2_1-z^2_2 + \eta(x_r + i z_1)(x^*_r + i z_2)\right]} dz_1 dz_2\right]\right\}_{\eta = - 2 |\xi_r|}.
\label{a3.6b}
\end{eqnarray}

Evaluating the integrals with the help of Fourier transform, we find
 \begin{eqnarray}
\fl \qquad  \quad G = \frac{4N^{2}_{r} |\xi_r|^2}{\pi}\left\{\frac{d^2}{d\eta^2}\left[\frac{1}{\sqrt{1 - \frac{\eta^2}{4}}}\exp{\left[\frac{\eta|x_r|^2 - \frac{\eta^2}{4}(x^2_r + x^{*^2}_r)}{1 - \frac{\eta^2}{4}}\right]}\right]\right\}_{\eta = - 2 |\xi_r|}.
\label{a3.7}
\end{eqnarray}

Now differentiating the terms inside the square bracket two times with respect to $\eta$ and then
substituting the expression for $N_r$ in the resultant equation and simplifying the latter we arrive at
\begin{eqnarray}
\fl \qquad G &=& \frac{1}{(1 - |\xi_r|^2)^2} \left[|\alpha_r|^4 (1 + |\xi_r|^2)^2 +  (\alpha^2 \xi^{*}_r + \alpha^{*^2}_{r} \xi_r)^2 + 4|\alpha_r|^2 |\xi_r|^2( 2 + |\xi_r|^2 )  \right. \nonumber \\
\fl \qquad & & \left. + |\xi_r|^2 (1 +   2|\xi_r|^2)+ ((2 |\alpha_r|^2(1 + |\xi_r|^2) + 1 + 5 |\xi_r|^2)(\alpha^2 \xi^{*}_r + \alpha^{*^2}_{r} \xi_r)\right].
\label{a3.8}
\end{eqnarray}

Substituting the equations (\ref{a3.8}) and (\ref{nrr}) in (\ref{a3.3}),
we get the following expression for $\langle \hat{n}^2_r \rangle$, that is
\begin{eqnarray}
\fl\;\; \langle \hat{n}^2_r \rangle &=& \frac{1}{ (1-|\xi_r|^2)^2 }\left[|\alpha_r|^4 (1 + |\xi_r|^2)^2 )+ |\xi_r|^2 (2 + |\xi_r|^2)  +(\xi^*_r \alpha^2_r + \xi_r \alpha^{*^2}_{r})^2 \right. \nonumber\\
\fl\;\;                            & &\left. \quad
 + (2 (1 + |\alpha_r|^2 )(1 + |\xi_r|^2) + 2 |\xi_r|^2) (\xi^*_r \alpha^2_r + \xi_r \alpha^{*^2}_{r})+  |\alpha_r|^2 (1 +  8|\xi_r|^2+ 3|\xi_r|^4 )\right].
\nonumber  \mbox{(\ref{nr2a})}
\end{eqnarray}
To numerically evaluate the Mandel's $Q$ parameter we use the expression (\ref{nr2a}).

\section{\bf Evaluation of Wigner function (\ref{wigf})}
\label{ap3}
Let us first evaluate the radial part in (\ref{wigf}). Separating the radial
part from  angular part we get
\begin{eqnarray}
\fl \qquad  W_r({\zeta_r}) = e^{-2|\zeta_r|^2} N^2_r \sum^{\infty}_{n = 0} \sum^{\infty}_{n' = 0} \frac{H_{n}(x_r)H_{n'}(x^*_r)}{n!}
(2 {\zeta^*_r})^{n - n'} \left(-\frac{\xi_r}{2}\right)^{n/2}\left(-\frac{\xi^{'}_r}{2}\right)^{n'/2} (-1)^{n'} \nonumber \\
\hspace{3cm}\times L^{n-n'}_{n'}(4 |\zeta_r|^2),
\label{a4.1}
\end{eqnarray}
where $x_r = \frac{\alpha_r \sqrt{1 - |\xi_r|^2}}{\sqrt{-2\xi_r}}$.

To start with we evaluate  the following sum appearing in (\ref{a4.1}), that is
\begin{eqnarray}
\fl \qquad\qquad\quad G_1 = \sum^{\infty}_{n = 0} \frac{H_{n}(x_r)}{n!} (2 \zeta^{*}_r)^n \left(\frac{-\xi_r}{2}\right)^{n/2} L^{n - n'}_{n'}(4 |\zeta|^2).
\label{a4.2}
\end{eqnarray}

Recalling the integral and summation representation of the Hermite and the
the associated Laguerre polynomials, namely
$H_{n}(z) = \frac{2^n}{\sqrt{\pi}}\int^{-\infty}_{\infty} e^{-t^2} (z + i t)^n dt$ and
$\sum^{\infty}_{n = 0} \frac{L^{k + \lambda}_n(z) w^k}{k!} = e^w L^{\lambda}_{n} (z - w)$ \cite{book},
we can reduce equation (\ref{a4.2}) to the form
\begin{eqnarray}
\fl \;G_1 = \frac{1}{\sqrt{\pi}}\nonumber \\
\fl \qquad \;\;  \times \int^{\infty}_{-\infty} \exp{\left[-t^2 + 2 (x_r + i t) \zeta^*_r \sqrt{- 2 \xi_r} \right]} L^{-n'}_{n'}\left(4 |\zeta_r|^2 - 2
\sqrt{-2 \xi_r} \zeta^{*}_r (x_r + i t)\right) dt.
\label{a4.3}
\end{eqnarray}

This expression can further be simplified by recalling yet another identity, that is  $L^{-n}_n(x) = \frac{(-x)^n}{n!}$.
With this simplification, equation (\ref{a4.3}) reshapes into
\begin{eqnarray}
\fl \;G_1 = \frac{(-1)^{n'}}{n'!\sqrt{\pi}} \nonumber \\
\fl \qquad \quad \times \int^{\infty}_{-\infty} \exp{\left[-t^2 + 2 (x_r + i t) \zeta^*_r \sqrt{- 2 \xi_r} \right]}
\left(4 |\zeta_r|^2 - 2\sqrt{-2 \xi_r} \zeta^{*}_r (x_r + i t)\right)^{n'} dt.
\label{a4.3b}
\end{eqnarray}

The second term in (\ref{a4.3b}) can also  be written as an exponential function through the relation
$\left(4 |\zeta_r|^2 - 2\sqrt{-2 \xi_r} \zeta^{*}_r (x_r + i t)\right)^{n'} = \left\{\frac{d^{n'}}{dp^{n'}} \left[\exp{\left[p(4|\zeta_r|^2 - 2 (x_r + i t) \sqrt{-2 \xi_r} \zeta^*_r)\right]}\right]\right\}_{p = 0}$. By doing so we find
\begin{eqnarray}
\fl \qquad  \qquad  \quad  G_1 = \frac{(-1)^{n'}}{n'!\sqrt{\pi}}\left\{\frac{d^{n'}}{dp^{n'}}\left[\exp{\left[4 p |\zeta_r|^2 + 2 (1-p)
\sqrt{-2\xi_r} \zeta^{*}_r x_r \right]} \right. \right. \nonumber \\
\qquad   \qquad  \quad \quad   \times \left. \left.\int^{\infty}_{-\infty} \exp{\left[-t^2 + 2 i (1-p) \sqrt{- 2 \xi_r} \zeta^*_r t\right]} dt \right]\right\}_{p =0}.
\label{a4.4}
\end{eqnarray}
Now evaluating the integral using the relation $\int^{\infty}_{-\infty} e^{i w x} e^{-x^2} dx = \sqrt{\pi} e^{-\frac{w^{2}}{4}}$,
we obtain
\begin{eqnarray}
\fl \;\;\;  \quad  G_1 = \frac{(-1)^{n'}}{n'!}\left\{\frac{d^{n'}}{dp^{n'}}\left[\exp{\left[4 p |\zeta_r|^2 + 2 (1-p)
\sqrt{-2\xi_r} \zeta^{*}_r x_r + 2 (1 - p)^2 \xi_r \zeta^{*^2}_r \right]}  \right]\right\}_{p =0}.
\label{a4.4b}
\end{eqnarray}

To proceed further we introduce the transformation, $\theta = \sqrt{-2\xi_r} \zeta^*_r (1 -p)$, in (\ref{a4.4b}). In this new
variable $\theta$, equation (\ref{a4.4b}) reads
\begin{eqnarray}
\fl \qquad \quad \quad  G_1 = \frac{(\sqrt{-2\xi_r} \zeta^*_r)^{n'}}{n'!}\left\{\exp{\left[4 |\zeta_r|^2
            + \left(x_r + \sqrt{\frac{-2}{\xi_r}} \zeta_r\right)^2\right]} \right. \nonumber \\
\fl \qquad \qquad \qquad \qquad  \qquad  \times \left.     \frac{d^{n'}}{d{\theta}^{n'}}\left[\exp{\left[ -\left(\theta -x_r
            - \sqrt{\frac{-2}{\xi_r}} \zeta_r\right)^2\right]}\right] \right\}_{\theta = \sqrt{-2 \xi_r} \zeta^*_r}.
\label{a4.5}
\end{eqnarray}
Expressing the differential part appearig in (\ref{a4.5}) in terms of Hermite polynomials through Rodrigues formula,
$H_{n}(x) = (-1)^n e^{x^2}\frac{d^{n}}{dx^n} e^{-x^2}$ \cite{book}, we find
\begin{eqnarray}
\fl \quad G_1 = \frac{(-\sqrt{-2 \xi_r} \zeta^{*}_r)^{n'}}{n'!} \exp{\left[2 \xi_r \zeta^{*^2}_r +  2 \sqrt{- 2 \xi_r}x_r
\zeta^{*}_r \right]} H_{n'}\left[\sqrt{- 2 \xi_r} \zeta^*_r - \sqrt{\frac{-2}{\xi_r}}\zeta_r - x_r\right]. \qquad
\label{a4.5b}
\end{eqnarray}
Substituting (\ref{a4.5b}) in the  Wigner function (\ref{a4.1}), we get
\begin{eqnarray}
\fl \; W_r = N^{2}_r\exp{\left[ - 2 |\zeta_r|^2 + 2 \xi_r \zeta^{*^2}_r +  2 \sqrt{- 2 \xi_r}x_r
\zeta^{*}_r \right]} \sum^{\infty}_{n' = 0} \frac{H_{n'}(x^*_r) H_{n'}(\tilde{\theta})}{n'!} \left(\frac{- |\xi_r|}{2}\right)^{n'},
\label{a4.6}
\end{eqnarray}
where $\tilde{\theta} =  \sqrt{- 2 \xi_r} \zeta^*_r- \sqrt{\frac{-2}{\xi_r}}\zeta_r - x_r$.

The  summation in (\ref{a4.6}) can be evaluated through the identity \cite{book}
\begin{eqnarray}
\fl \qquad \quad  \sum^{\infty}_{n = 0} \frac{H_{n}(x) H_{n}(y) w^n}{n!} = \frac{1}{\sqrt{1 - 4w^2}} \exp{\left[ \frac{4 w^2 (x^2 + y^2) - 4 w x y}{4 w^2 -1}\right]}, \; |w| < \frac{1}{2}.
\end{eqnarray}
On replacing this result in (\ref{a4.6}), we obtain
\begin{eqnarray}
\fl \quad W_r = N^{2}_r \exp{\left[-2\xi \zeta^{*^2}_r + 2 \sqrt{-2 \xi_r} \zeta^{*}_r -  2 |\zeta_r|^2 + \
\frac{|\xi^2_r|}{|\xi_r|^2 - 1}|\left( x^2_r + x^{*^2}_r + 4 |\zeta_r|^2 \right. \right. } \nonumber \\
\quad {\left. \left.- 2 \xi_r \zeta^{*^2}_r - \frac{2}{\xi_r} \zeta^2_r
  + 2 \sqrt{\frac{-2}{\xi_r}} \zeta_r - 2 \sqrt{-2 \xi_r} x_r \zeta^*_r\right) + \frac{2 |\xi_r|}{|\xi_r|^2 -1}\left( - |x_r|^2 \right. \right. } \nonumber \\
 \qquad{\left. \left.- \frac{-2}{\xi_r} x^*_r + \sqrt{-2 \xi_r} \zeta^*_r x^*_r \right)\right]}.
\label{a4.7a}
\end{eqnarray}

The final task is to substitute the normalization constant, (\ref{nr}), in (\ref{a4.7a})
and simplify the resultant expression with
$x_r = \frac{\alpha_r \sqrt{1 - |\xi_r|^2}}{\sqrt{-2 \xi_r}}$. On completing this task we arrive at
the following expression for the radial part, that is
\begin{eqnarray}
\fl \quad W_r &=& \exp{\left[\frac{-2|\zeta_r|^2 (|\xi_r|^2 + 1) + 2 (\xi_r \zeta^{*^2}_r + \xi^*_r \zeta^2_r)}{1 - |\xi_r|^2} - 2 |\alpha_r|^2   + \frac{2(\alpha_r \zeta^{*}_r + \alpha^{*}_r \zeta_r)}{\sqrt{1 - |\xi_r|^2}} \right. }\nonumber \\
\fl \quad       & & \quad \quad {\left. - \frac{2  (\xi^{*}_r\alpha_r\zeta_r + \xi_r\alpha^*_{r} \zeta^{*}_r)}{\sqrt{1-|\xi_r|^2}}\right]}.
\label{a4.7}
\end{eqnarray}

In a similar way one can also evaluate the other two modes present in the Wigner function. Our result shows that
(since the procedure is repetitive, in the following, we give only the final form of the expression)
\begin{eqnarray}
\fl \qquad \qquad W_{\pm}&=& \exp{\left[\frac{-2|\zeta_{\pm}|^2 (|\xi_{\pm}|^2 + 1) + 2 (\xi_{\pm} \zeta^{*^2}_{\pm} + \xi^*_{\pm} \zeta^2_{\pm})}{1 - |\xi_{\pm}|^2} - 2 |\alpha_{\pm}|^2  \right. } \nonumber \\
\fl \qquad\quad & & \qquad \qquad \qquad{\left.+ \frac{2(\alpha_{\pm} \zeta^{*}_{\pm} + \alpha^{*}_{\pm} \zeta_{\pm})}{\sqrt{1 - |\xi_{\pm}|^2}}
 - \frac{2  (\xi^{*}_{\pm}\alpha_{\pm}\zeta_{\pm} + \xi_{\pm}\alpha^*_{\pm} \zeta^{*}_{\pm})}{\sqrt{1-|\xi_{\pm}|^2}}\right]}.
\label{a4.8}
\end{eqnarray}

As a result, we obtain the following form for the Wigner function for the three dimensional generalized
isotonic oscillator, that is
\begin{eqnarray}
\fl \quad W(\{\zeta_r, \zeta_{+}, \zeta_{-}\}) &=& W_{r} \times W_{\pm} \nonumber \\
\fl \quad                                      &=& \Pi^{3}_{j=1}\exp{\left[\frac{-2|\zeta_{j}|^2 (|\xi_{j}|^2 + 1) + 2 (\xi_{j} \zeta^{*^2}_{j} + \xi^*_j \zeta^2_{j})}{1 - |\xi_{j}|^2} - 2 |\alpha_{j}|^2 \right. }\nonumber \\
 \fl \quad       & & \qquad \qquad \qquad{\left.+ \frac{2(\alpha_j \zeta^{*}_j + \alpha^{*}_j \zeta_j)}{\sqrt{1 - |\xi_j|^2}}
 - \frac{2  (\xi^{*}_j\alpha_r\zeta_j+ \xi_r\alpha^*_{j} \zeta^{*}_j)}{\sqrt{1-|\xi_j|^2}}\right]}.
\label{a4.9}
\end{eqnarray}

We use this result to investigate the Wigner function numerically.
\end{appendix}

\end{document}